\documentclass[pre,aps,floats,twocolumn,superscriptaddress,floatfix]{revtex4}
\usepackage{graphicx,amssymb,amsmath,ifthen,rotating}
\usepackage[active]{srcltx}
\usepackage{tikz}
\usepackage{verbatim}
\begin{document}
\title{Jammed disks of two sizes in a channel:
segregation driven by steric forces}  
  \author{Dan Liu}
\affiliation{
  Department of Physics,
  University of Hartford,
  West Hartford, CT 06117, USA}
      \author{Michael Karbach}
\affiliation{
Fachgruppe Physik,
  Bergische Universit{\"{a}}t Wuppertal,
  D-42097 Wuppertal, Germany}
\author{Gerhard M{\"{u}}ller}
\affiliation{
  Department of Physics,
  University of Rhode Island,
  Kingston RI 02881, USA}
\begin{abstract}
Disks of two sizes are confined to a long and narrow channel. 
The axis and the plane of the channel are horizontal.
The channel is closed off by pistons that freeze jammed microstates out of loose disk configurations, agitated randomly at calibrated intensity and subject to moderate pressure.
Disk sizes and channel width are such that under jamming no disks remain loose and all disks touch one wall.
The protocol permits disks to move past each other prior to jamming, which facilitates randomness in the sequence of large and small disks.
We present exact results for the characterization of jammed macrostates including volume and entropy for given fractions of small and large disks as functions of energy parameters which depend on the jamming protocol.
Our analysis divides the disk sequence of jammed microstates into overlapping tiles out of which we construct 17 species of statistically interacting quasiparticles. 
Jammed macrostates then depend on the fractions of small and large disks and on a dimensionless control parameter inferred from measures for expansion work against the pistons and intensity of random agitations.
Two models are introduced for comparison of key technical aspects: one model emphasizes symmetry and the other mechanical stability.
We distinguish regimes for the energy parameters that either enhance or suppress mixing of disk sizes in jammed macrostates. 
The latter case, if realizable, is a manifestation of grain segregation driven by steric forces alone, without directional bias.
\end{abstract}
\maketitle

%
\section{Introduction}\label{sec:intro}
%
Randomness in jammed granular matter is a multifaceted phenomenon \cite{JN96, DeGe99, MBE05, Meht10, TS10, Xu11, BMHM18, BB19}.
Jammed monodisperse grains exhibit configurational randomness, quantified by configurational entropy and average density.
In the presence of external fields, such as provided by gravity or centrifugation, the configurational entropy and average density turn into scalar fields.
Polydisperse granular matter offers additional sources of randomness due to variety in weight, size, and shape of the grains.
While different forms of randomness are neither fully separable nor additive, they each tend to leave distinctive signatures in jammed macrostates, identifiable in experiments \cite{KFL+95, MJS00, TPC+01, LDDB06, SNRS07, MSLB07, GBOS09, ZXC+09, ZMSB10, NJVB11, ENW12, PD13, BKD19} and, in many cases, reproducible in simulation studies \cite{OLLN01, OLLN02, XBO05, ZM05, UKW05, SDST06, DST06, GBO06, OSN06, SVE+07, SWM08, Head09, DW09, JCM+10, CCC10, JM10, SOS11, XFL11, OH11, CPNC11, PCC12, CCPZ12, HB12, LFS+18}.

Grain segregation by size, weight, or shape is a common observation in polydisperse granular matter.
Depending on the context, unmixing is desirable or undesirable. 
In either case, we wish to understand how different agents enhance or suppress segregation according to size.
The most common causes of segregation are a gravitational field, centrifugal forces, and container walls of particular shapes. 
A commonality of these agents is that each favors, in one way or another, a direction of segregation \cite{WH22, WLG+23, ABS22, DPU+22, PHBJ22, ONV+17, Zacc22, ACZ+23, Zacc25}. 

These considerations raise two important questions: (i) What is the opposite of a segregation tendency in granular matter? 
Is it a tendency toward (low-entropy) uniformity or a tendency toward (high-entropy) randomness?
(ii) Might there be scenarios in which grain segregation is associated with spontaneous symmetry breaking as is known in phase transitions of particles sufficiently small to be impacted by thermal fluctuations?
This work intends to throw new light on both questions.

It may be difficult to recognize spontaneous symmetry breaking as the cause of grain segregation in experiments or in simulation studies even if such a cause is in operation.
A theoretical model that predicts the effect and describes scenarios where it could play out, would, therefore, be quite helpful.
Alas, in a field dominated by experiments and simulations as is the case in granular matter research, theoretical approaches face daunting challenges.
The adaptation of standard tools from statistical mechanics and fluid dynamics to granular matter is tricky and limited in many ways.

The best chances for inroads are likely to present themselves to studies of polydisperse grains in narrow channels, where caging effects permit a systematic characterization of jammed microstates \cite{LFS+18, RVDB22, JEW23, LWH+22}.
Significant advances in the study of grains thus confined have employed a variety of strategies (mostly for monodisperse systems) \cite{LM07, LM10, BS06, AB09, BA11, YAB12, AYB13, janac1, janac2}.
The results of such projects are illuminating and complement simulation data for comparable situations in the sense that the same or similar outcomes are detected through different filters.

A recent study with focus on alternating sequences of jammed disks of two sizes and weights in a narrow channel \cite{LM19, janac3} was, in significant aspects, preparatory in relation to this work.
It dealt with jammed configurations of alternating sequences of disks that come in two sizes. 
It employed a methodology, previously developed for monodisperse systems \cite{janac1, janac2}, which treats jammed disk configurations as statistically interacting particles activated from a reference state of choice. 
The analysis is rigorous within the framework of configurational statistics applied to granular matter.

The goal of studying situations that make grain segregation possible, as endeavored in the work reported here, requires two major extensions in the approach developed thus far.
(i) The jamming protocol must be modified such that it permits the randomization of disks of different sizes.
(ii) The concept of grandcanonical ensemble must be adapted to granular statistics for counting purposes, which requires the use of a quantity akin to a chemical potential \cite{janac3, GZB23}.
With these extensions we can demonstrate that size segregation is possible as a result of steric forces alone and with no directional bias caused by external fields or the profile of walls.

We first set the stage with specifications regarding the geometry of jammed disk placements in the channel (Sec.~\ref{sec:geom}), the combinatorics of disk configurations in jammed macrostates (Sec.~\ref{sec:comb}), and the energetics of disks prior to jamming (Sec.~\ref{sec:ener}).
Next we outline the methodology (Sec.~\ref{sec:stat}) for working out the exact configurational statistical analysis applied to a scenario of maximum symmetry (Sec.~\ref{sec:scen1}).
For the (undetermined) energy parameters we identify two regimes, potentially realizable in different jamming protocols, which lead to manifestations of either \emph{size segregation} or \emph{uniform size mixing} -- two distinct types of long-range order.
For energy parameters on the border between the two regimes, no ordering tendency prevails and \emph{size randomness} characterizes the macrostate.
We demonstrate the robustness of our findings by comparing them with those of a scenario that is less symmetric but eliminates jammed configurations that are not completely stable  (Sec.~\ref{sec:scen2}).

%
\section{Disks, Tiles, and Particles}\label{sec:geom}
%
Disks with diameters $\sigma_\mathrm{L}\geq\sigma_\mathrm{S}$ in arbitrary sequence are being jammed in a channel of width $H$.
Every jammed disk has three points of contact with an adjacent disk or a wall. 
The constraints,  
\begin{align}\label{eq:1} 
1\leq \frac{\sigma_\mathrm{L}}{\sigma_\mathrm{S}}
<\frac{H}{\sigma_\mathrm{S}}<
1+\sqrt{3/4},
\end{align}
guarantee that every jammed disk touches one wall and two adjacent disk. 
There are no loose disks.
Mechanical stability requires that the angle between any two contacts of a disk falls below $180^\circ$. 
Configurations in which one angle is $180^\circ$ are called marginally stable.

All jammed microstates can be assembled from 16 tiles composed of two adjacent disks similar to domino pieces with one disk overlapping (Table~\ref{tab:t1}).
Adding a tile to an already existing string of tiles must satisfy two successor rules:
The tile added must (i) match the pattern regarding size and position of the overlapping disk and (ii) maintain mechanical stability under jamming forces.
Each tile has one of six distinct volumes (Table ~\ref{tab:t2}).

\begin{table}[htb]
  \caption{Distinct tiles that constitute jammed microstates of arbitrary disk sequences subject to the constraints (\ref{eq:1}). 
The ``ID'' tile must be followed one of the ``rule'' tiles. 
The motifs shown pertain to $\sigma_\mathrm{L}=2$, $\sigma_\mathrm{S}=1.4$, $H=2.5$. 
Rule entries in square-brackets permit jammed configurations of marginal stability as explained in the text.}\label{tab:t1}
\begin{center}
\begin{tabular}{cccc|cccc} \hline\hline
motif & ~ID~ & rule & vol.~ & ~motif~ & ~ID~ & rule & vol.  \\ \hline \rule[-2mm]{0mm}{8mm}
\begin{tikzpicture} [scale=0.2]
\draw (0.0,0.0) -- (3.94,0.0) -- (3.94,2.5) -- (0.0,2.5) -- (0,0);
\filldraw [fill=gray, draw=black] (1.0,1.5) circle (1.0);
\filldraw [fill=gray, draw=black] (2.94,1.0) circle (1.0);
\end{tikzpicture}
& 1 & $2,6,10,14$ & $V_\mathrm{a}$ &   
\begin{tikzpicture} [scale=0.2]
\draw (0.0,0.0) -- (4.0,0.0) -- (4.0,2.5) -- (0.0,2.5) -- (0,0);
\filldraw [fill=gray, draw=black] (1,1.5) circle (1.0);
\filldraw [fill=gray, draw=black] (3.0,1.5) circle (1.0);
\end{tikzpicture}
& 9 & $1,5,[9]$ & $V_\mathrm{d}$  \\ \rule[-2mm]{0mm}{6mm}

\begin{tikzpicture} [scale=0.2]
\draw (0.0,0.0) -- (3.94,0.0) -- (3.94,2.5) -- (0.0,2.5) -- (0,0);
\filldraw [fill=gray, draw=black] (1,1) circle (1.0);
\filldraw [fill=gray, draw=black] (2.94,1.5) circle (1.0);
\end{tikzpicture}
& 2 & $1,5,9,13$ & $V_\mathrm{a}$ &  
\begin{tikzpicture} [scale=0.2]
\draw (0.0,0.0) -- (4.0,0.0) -- (4.0,2.5) -- (0.0,2.5) -- (0,0);
\filldraw [fill=gray, draw=black] (1.0,1.0) circle (1.0);
\filldraw [fill=gray, draw=black] (3.0,1.0) circle (1.0);
\end{tikzpicture}
& 10 & $2,6,[10]$ & $V_\mathrm{d}$  \\ \rule[-2mm]{0mm}{6mm}

\begin{tikzpicture} [scale=0.2]
\draw (0.0,0.0) -- (2.27,0.0) -- (2.27,2.5) -- (0.0,2.5) -- (0,0);
\filldraw [fill=gray, draw=black] (0.7,1.8) circle (0.7);
\filldraw [fill=gray, draw=black] (1.57,0.7) circle (0.7);
\end{tikzpicture}
& 3 & $4,8,12,16$ & $V_\mathrm{b}$ & 
\begin{tikzpicture} [scale=0.2]
\draw (0.0,0.0) -- (2.8,0.0) -- (2.8,2.5) -- (0.0,2.5) -- (0,0);
\filldraw [fill=gray, draw=black] (0.7,1.8) circle (0.7);
\filldraw [fill=gray, draw=black] (2.1,1.8) circle (0.7);
\end{tikzpicture}
& 11 & $3,7,15,[11]$ & $V_\mathrm{e}$  \\ \rule[-2mm]{0mm}{6mm}

\begin{tikzpicture} [scale=0.2]
\draw (0.0,0.0) -- (2.27,0.0) -- (2.27,2.5) -- (0.0,2.5) -- (0,0);
\filldraw [fill=gray, draw=black] (0.7,0.7) circle (0.7);
\filldraw [fill=gray, draw=black] (1.57,1.8) circle (0.7);
\end{tikzpicture}
& 4 & $3,7,11,15$ & $V_\mathrm{b}$ & 
\begin{tikzpicture} [scale=0.2]
\draw (0.0,0.0) -- (2.8,0.0) -- (2.8,2.5) -- (0.0,2.5) -- (0,0);
\filldraw [fill=gray, draw=black] (0.7,0.7) circle (0.7);
\filldraw [fill=gray, draw=black] (2.1,0.7) circle (0.7);
\end{tikzpicture}
& 12 & $4,8,16,[12]$ & $V_\mathrm{e}$ \\ \rule[-2mm]{0mm}{6mm}

\begin{tikzpicture} [scale=0.2]
\draw (0.0,0.0) -- (3.2,0.0) -- (3.2,2.5) -- (0.0,2.5) -- (0,0);
\filldraw [fill=gray, draw=black] (1,1.5) circle (1.0);
\filldraw [fill=gray, draw=black] (2.5,0.7) circle (0.7);
\end{tikzpicture}
& 5 & $4,8,12,16$ & $V_\mathrm{c}$ &   
\begin{tikzpicture} [scale=0.2]
\draw (0.0,0.0) -- (3.37,0.0) -- (3.37,2.5) -- (0.0,2.5) -- (0,0);
\filldraw [fill=gray, draw=black] (1.0,1.5) circle (1.0);
\filldraw [fill=gray, draw=black] (2.67,1.8) circle (0.7);
\end{tikzpicture}
& 13 & $3,7,15,11$ & $V_\mathrm{f}$  \\ \rule[-2mm]{0mm}{6mm}

\begin{tikzpicture} [scale=0.2]
\draw (0.0,0.0) -- (3.2,0.0) -- (3.2,2.5) -- (0.0,2.5) -- (0,0);
\filldraw [fill=gray, draw=black] (1,1) circle (1.0);
\filldraw [fill=gray, draw=black] (2.5,1.8) circle (0.7);
\end{tikzpicture}
& 6 & $3,7,11,15$ & $V_\mathrm{c}$ &  
\begin{tikzpicture} [scale=0.2]
\draw (0.0,0.0) -- (3.37,0.0) -- (3.37,2.5) -- (0.0,2.5) -- (0,0);
\filldraw [fill=gray, draw=black] (1.0,1.0) circle (1.0);
\filldraw [fill=gray, draw=black] (2.677,0.7) circle (0.7);
\end{tikzpicture}
& 14 & $4,8,16,12$ & $V_\mathrm{f}$  \\ \rule[-2mm]{0mm}{6mm}

\begin{tikzpicture} [scale=0.2]
\draw (0.0,0.0) -- (3.2,0.0) -- (3.2,2.5) -- (0.0,2.5) -- (0,0);
\filldraw [fill=gray, draw=black] (0.7,1.8) circle (0.7);
\filldraw [fill=gray, draw=black] (2.2,1.0) circle (1.0);
\end{tikzpicture}
& 7 & $2,6,10,14$ & $V_\mathrm{c}$ & 
\begin{tikzpicture} [scale=0.2]
\draw (0.0,0.0) -- (3.37,0.0) -- (3.37,2.5) -- (0.0,2.5) -- (0,0);
\filldraw [fill=gray, draw=black] (0.7,1.8) circle (0.7);
\filldraw [fill=gray, draw=black] (2.37,1.5) circle (1.0);
\end{tikzpicture}
& 15 & $1,5$& $V_\mathrm{f}$  \\ \rule[-2mm]{0mm}{6mm}

\begin{tikzpicture} [scale=0.2]
\draw (0.0,0.0) -- (3.2,0.0) -- (3.2,2.5) -- (0.0,2.5) -- (0,0);
\filldraw [fill=gray, draw=black] (0.7,0.7) circle (0.7);
\filldraw [fill=gray, draw=black] (2.2,1.5) circle (1.0);
\end{tikzpicture}
& 8 & $1,5,9,13$ & $V_\mathrm{c}$ & 
\begin{tikzpicture} [scale=0.2]
\draw (0.0,0.0) -- (3.37,0.0) -- (3.37,2.5) -- (0.0,2.5) -- (0,0);
\filldraw [fill=gray, draw=black] (0.7,0.7) circle (0.7);
\filldraw [fill=gray, draw=black] (2.37,1.0) circle (1.0);
\end{tikzpicture}
& 16 & $2,6$ & $V_\mathrm{f}$ 
\\ \hline\hline
\end{tabular}
\end{center}
\end{table}

\begin{table}[b]
  \caption{Volume of tiles in a channel has unit cross sections. The bracketed portion overlaps in successive tiles and does not contribute the excess volume of quasiparticles.
  The numerical values pertain to $\sigma_\mathrm{L}=2$, $\sigma_\mathrm{S}=1.4$, $H=2.5$. The disks have unit height, matching the channel width in that direction.}\label{tab:t2}
\begin{center}
\begin{tabular}{l|l|l} \hline\hline
 & ~$V=[~~]+\tilde{V}$ & ~$\tilde{V}$  \\ \hline \rule[-2mm]{0mm}{6mm}
$V_\mathrm{a}$~ & ~$[\sigma_\mathrm{L}]+\sqrt{H(2\sigma_\mathrm{L}-H)}$ & ~1.963 
\\ \rule[-2mm]{0mm}{6mm}
$V_\mathrm{b}$~ & ~$[\sigma_\mathrm{S}]+\sqrt{H(2\sigma_\mathrm{S}-H)}$ & ~0.886 
\\ \rule[-2mm]{0mm}{6mm}
$V_\mathrm{c}$~ & ~$[\frac{1}{2}(\sigma_\mathrm{L}+\sigma_\mathrm{S})]
+\sqrt{H(\sigma_\mathrm{L}+\sigma_\mathrm{S}-H)}$~~ & ~1.5 
\\ \rule[-2mm]{0mm}{6mm}
$V_\mathrm{d}$~ & ~$[\sigma_\mathrm{L}]+\sigma_\mathrm{L}$ & ~2.0 
\\ \rule[-2mm]{0mm}{6mm}
$V_\mathrm{e}$~ & ~$[\sigma_\mathrm{S}]+\sigma_\mathrm{S}$ & ~1.4 
\\ \rule[-2mm]{0mm}{6mm}
$V_\mathrm{f}$~ & ~$[\frac{1}{2}(\sigma_\mathrm{L}+\sigma_\mathrm{S})]
+\sqrt{\sigma_\mathrm{L}\sigma_\mathrm{S}}$ & ~1.673 
\\ \hline\hline
\end{tabular}
\end{center}
\end{table}

Jammed-disk microstates are described as configurations of statistically interacting particles generated from a reference state (pseudo-vacuum) of choice.
We have selected two alternative reference states, one containing only large disks and the other only small disks.
The former is constructed from tiles 1 and 2 exclusively:
\begin{equation}\label{eq:2} 
\mathsf{pv_L}=\mathsf{1212\cdots 1}\qquad 
\begin{tikzpicture} [scale=0.2]
\draw (0,0) -- (10,0.0);
\draw (0,0) -- (0,2.5);
\draw (0,2.5) -- (10,2.5);
\filldraw [fill=gray, draw=black] (1.0,1.5) circle (1.0);
\filldraw [fill=gray, draw=black] (2.94,1.0) circle (1.0);
\filldraw [fill=gray, draw=black] (4.88,1.5) circle (1.0);
\filldraw [fill=gray, draw=black] (6.82,1.0) circle (1.0);
\filldraw [fill=gray, draw=black] (8.74,1.5) circle (1.0);
\filldraw [fill=gray, draw=black] (11,1.25) circle (0.1);
\filldraw [fill=gray, draw=black] (12,1.25) circle (0.1);
\filldraw [fill=gray, draw=black] (13,1.25) circle (0.1);
\end{tikzpicture}\hspace{2mm}
\begin{tikzpicture} [scale=0.2]
\draw (0.0,0.0) -- (4.14,0.0) -- (4.14,2.5) -- (0.0,2.5);
\filldraw [fill=gray, draw=black] (1.2,1.5) circle (1.0);
\filldraw [fill=gray, draw=black] (3.14,1.0) circle (1.0);
\end{tikzpicture}
\end{equation}
All other jammed microstates can be generated by the activation of quasiparticles from this reference state. 
We have identified 17 species of particles that serve this purpose (Table~\ref{tab:t3}).
The taxonomy of Ref.~\cite{copic} introduces categories of particle species adopted here.
The \emph{compacts} $m=1,2$ and \emph{hosts} $m=3,\ldots,14$ modify the pseudo-vacuum (\ref{eq:2}), whereas the \emph{tags} $m=15,16,17$ modify any one of the hosts. 
All jammed configurations thus generated are stable with two exception.
Particles from species $m=5,6$ are unstable in some configurations.
Every occurrence of marginal stability is associated with particles from species $m=1,2,15,16$.
Ajacent particles or adjacent elements of vacuum share one disk.

If we interchange the roles of large and small disks, we produce a different set of quasiparticles (Table~\ref{tab:t7}). 
The pseudo-vacuum for these particles is the tile sequence,
\begin{equation}\label{eq:36} 
\mathsf{pv_S}=\mathsf{3434\cdots 3}\qquad
\begin{tikzpicture} [scale=0.2]
\draw (0,0) -- (0,2.5);
\draw (0,0) -- (5.18,0);
\draw (0,2.5) -- (5.18,2.5);
\filldraw [fill=gray, draw=black] (0.7,1.8) circle (0.7);
\filldraw [fill=gray, draw=black] (1.57,0.7) circle (0.7);
\filldraw [fill=gray, draw=black] (2.44,1.8) circle (0.7);
\filldraw [fill=gray, draw=black] (3.31,0.7) circle (0.7);
\filldraw [fill=gray, draw=black] (4.18,1.8) circle (0.7);
\filldraw [fill=gray, draw=black] (6,1.25) circle (0.1);
\filldraw [fill=gray, draw=black] (7,1.25) circle (0.1);
\filldraw [fill=gray, draw=black] (8,1.25) circle (0.1);
\end{tikzpicture}\hspace{2mm}
\begin{tikzpicture} [scale=0.2]
\draw (0.0,0.0) -- (2.47,0.0) -- (2.47,2.5) -- (0.0,2.5);
\filldraw [fill=gray, draw=black] (0.9,1.8) circle (0.7);
\filldraw [fill=gray, draw=black] (1.77,0.7) circle (0.7);
\end{tikzpicture}
\end{equation}
representing the state with the highest number of disks per unit length overall.
The categories remain the same: compacts $m=1,2$, hosts $m=3,\ldots,14$ and tags $m=15,16,17$. 
Most particles are again stable.
Particles $5,6$ are unstable, whereas particles $1,2,15,16$ in some configurations are only marginally stable.
Ways to retain only the stable configurations will be implemented in Sec.~\ref{sec:scen2}.
We shall demonstrate that the main conclusions are not affected by these instances of marginal stability or instability.

Finally, particles $m=7,8,13,14,17$ have two distinct motifs in Tables~\ref{tab:t3} and \ref{tab:t7}. 
Wherever there is an open slot for one of these particles, exactly one of the two motifs will fit to make a mechanically stable configuration. The two motifs are not interchangeable. Since the two motifs have the same disk content and the same excess volume, they represent the same particle.

\begin{table}[t]
  \caption{Species of quasiparticles which can be activated (directly or indirectly) from the reference state (\ref{eq:2}). The motifs shown are for $\sigma_\mathrm{L}=2$, $\sigma_\mathrm{S}=1.4$, $H=2.5$. The excess volumes $\Delta V_m$ are relative to a segment of reference state $\mathsf{pv_L}$ with the same number of disks. Species $m=7,8,13,14,17$ have two distinct motifs.}\label{tab:t3}
\begin{center}
\begin{tabular}{rl|rl|l} \hline\hline
$m$~ & motif & ~$m$~ & motif & ~$\Delta V_m$ \\ \hline \rule[-2mm]{0mm}{8mm}
1~ & \begin{tikzpicture} [scale=0.2]
\draw (0.0,0.0) -- (4.0,0.0) -- (4.0,2.5) -- (0.0,2.5) -- (0,0);
\filldraw [fill=gray, draw=black] (1,1.5) circle (1.0);
\filldraw [fill=gray, draw=black] (3.0,1.5) circle (1.0);
\end{tikzpicture}
& 
2~ & \begin{tikzpicture} [scale=0.2]
\draw (0.0,0.0) -- (4.0,0.0) -- (4.0,2.5) -- (0.0,2.5) -- (0,0);
\filldraw [fill=gray, draw=black] (1.0,1.0) circle (1.0);
\filldraw [fill=gray, draw=black] (3.0,1.0) circle (1.0);
\end{tikzpicture}
& ~$\tilde{V}_\mathrm{d}-\tilde{V}_\mathrm{a}$
\\ \rule[-2mm]{0mm}{6mm}
3~ & \begin{tikzpicture} [scale=0.2]
\draw (0.0,0.0) -- (5.0,0.0) -- (5.0,2.5) -- (0.0,2.5) -- (0,0);
\filldraw [fill=gray, draw=black] (1,1.5) circle (1.0);
\filldraw [fill=gray, draw=black] (2.5,0.7) circle (0.7);
\filldraw [fill=gray, draw=black] (4.0,1.5) circle (1.0);
\end{tikzpicture}
& 
4~ & \begin{tikzpicture} [scale=0.2]
\draw (0.0,0.0) -- (5.0,0.0) -- (5.0,2.5) -- (0.0,2.5) -- (0,0);
\filldraw [fill=gray, draw=black] (1,1.0) circle (1.0);
\filldraw [fill=gray, draw=black] (2.5,1.8) circle (0.7);
\filldraw [fill=gray, draw=black] (4.0,1.0) circle (1.0);
\end{tikzpicture}
& ~$2\tilde{V}_\mathrm{c}-2\tilde{V}_\mathrm{a}$
\\ \rule[-2mm]{0mm}{6mm}
5~ & \begin{tikzpicture} [scale=0.2]
\draw (0.0,0.0) -- (5.4,0.0) -- (5.4,2.5) -- (0.0,2.5) -- (0,0);
\filldraw [fill=gray, draw=black] (1.0,1.5) circle (1.0);
\filldraw [fill=gray, draw=black] (2.7,1.8) circle (0.7);
\filldraw [fill=gray, draw=black] (4.4,1.5) circle (1.0);
\end{tikzpicture}
& 
6~ & \begin{tikzpicture} [scale=0.2]
\draw (0.0,0.0) -- (5.4,0.0) -- (5.4,2.5) -- (0.0,2.5) -- (0,0);
\filldraw [fill=gray, draw=black] (1.0,1.0) circle (1.0);
\filldraw [fill=gray, draw=black] (2.7,0.7) circle (0.7);
\filldraw [fill=gray, draw=black] (4.4,1.0) circle (1.0);
\end{tikzpicture}
& ~$2\tilde{V}_\mathrm{f}-2\tilde{V}_\mathrm{a}$
\\ \rule[-2mm]{0mm}{6mm}
7~ & \begin{tikzpicture} [scale=0.2]
\draw (0.0,0.0) -- (5.2,0.0) -- (5.2,2.5) -- (0.0,2.5) -- (0,0);
\filldraw [fill=gray, draw=black] (1.0,1.0) circle (1.0);
\filldraw [fill=gray, draw=black] (2.5,1.8) circle (0.7);
\filldraw [fill=gray, draw=black] (4.2,1.5) circle (1.0);
\end{tikzpicture}
&  
7~ & \begin{tikzpicture} [scale=0.2]
\draw (0.0,0.0) -- (5.2,0.0) -- (5.2,2.5) -- (0.0,2.5) -- (0,0);
\filldraw [fill=gray, draw=black] (1.0,1.5) circle (1.0);
\filldraw [fill=gray, draw=black] (2.7,1.8) circle (0.7);
\filldraw [fill=gray, draw=black] (4.2,1.0) circle (1.0);
\end{tikzpicture}
& ~$\tilde{V}_\mathrm{c}+\tilde{V}_\mathrm{f}-2\tilde{V}_\mathrm{a}$
\\ \rule[-2mm]{0mm}{6mm}
8~ & \begin{tikzpicture} [scale=0.2]
\draw (0.0,0.0) -- (5.2,0.0) -- (5.2,2.5) -- (0.0,2.5) -- (0,0);
\filldraw [fill=gray, draw=black] (1.0,1.5) circle (1.0);
\filldraw [fill=gray, draw=black] (2.5,0.7) circle (0.7);
\filldraw [fill=gray, draw=black] (4.2,1.0) circle (1.0);
\end{tikzpicture}
&  
8~ & \begin{tikzpicture} [scale=0.2]
\draw (0.0,0.0) -- (5.2,0.0) -- (5.2,2.5) -- (0.0,2.5) -- (0,0);
\filldraw [fill=gray, draw=black] (1.0,1.0) circle (1.0);
\filldraw [fill=gray, draw=black] (2.7,0.7) circle (0.7);
\filldraw [fill=gray, draw=black] (4.2,1.5) circle (1.0);
\end{tikzpicture}
& ~$\tilde{V}_\mathrm{c}+\tilde{V}_\mathrm{f}-2\tilde{V}_\mathrm{a}$
\\ \rule[-2mm]{0mm}{6mm}
9~ & \begin{tikzpicture} [scale=0.2]
\draw (0.0,0.0) -- (6.07,0.0) -- (6.07,2.5) -- (0.0,2.5) -- (0,0);
\filldraw [fill=gray, draw=black] (1.0,1.5) circle (1.0);
\filldraw [fill=gray, draw=black] (2.7,1.8) circle (0.7);
\filldraw [fill=gray, draw=black] (3.57,0.7) circle (0.7);
\filldraw [fill=gray, draw=black] (5.07,1.5) circle (1.0);
\end{tikzpicture}
&  
10~ & \begin{tikzpicture} [scale=0.2]
\draw (0.0,0.0) -- (6.07,0.0) -- (6.07,2.5) -- (0.0,2.5) -- (0,0);
\filldraw [fill=gray, draw=black] (1.0,1.0) circle (1.0);
\filldraw [fill=gray, draw=black] (2.5,1.8) circle (0.7);
\filldraw [fill=gray, draw=black] (3.37,0.7) circle (0.7);
\filldraw [fill=gray, draw=black] (5.07,1.0) circle (1.0);
\end{tikzpicture}
&  ~$\tilde{V}_\mathrm{b}+\tilde{V}_\mathrm{c}+\tilde{V}_\mathrm{f}-3\tilde{V}_\mathrm{a}$
\\ \rule[-2mm]{0mm}{6mm}
11~ & \begin{tikzpicture} [scale=0.2]
\draw (0.0,0.0) -- (6.07,0.0) -- (6.07,2.5) -- (0.0,2.5) -- (0,0);
\filldraw [fill=gray, draw=black] (1.0,1.5) circle (1.0);
\filldraw [fill=gray, draw=black] (2.5,0.7) circle (0.7);
\filldraw [fill=gray, draw=black] (3.37,1.8) circle (0.7);
\filldraw [fill=gray, draw=black] (5.07,1.5) circle (1.0);
\end{tikzpicture}
&  
12~ & \begin{tikzpicture} [scale=0.2]
\draw (0.0,0.0) -- (6.07,0.0) -- (6.07,2.5) -- (0.0,2.5) -- (0,0);
\filldraw [fill=gray, draw=black] (1.0,1.0) circle (1.0);
\filldraw [fill=gray, draw=black] (2.7,0.7) circle (0.7);
\filldraw [fill=gray, draw=black] (3.57,1.8) circle (0.7);
\filldraw [fill=gray, draw=black] (5.07,1.0) circle (1.0);
\end{tikzpicture}
&  ~$\tilde{V}_\mathrm{b}+\tilde{V}_\mathrm{c}+\tilde{V}_\mathrm{f}-3\tilde{V}_\mathrm{a}$
\\ \rule[-2mm]{0mm}{6mm}
13~ & \begin{tikzpicture} [scale=0.2]
\draw (0.0,0.0) -- (6.27,0.0) -- (6.27,2.5) -- (0.0,2.5) -- (0,0);
\filldraw [fill=gray, draw=black] (1.0,1.5) circle (1.0);
\filldraw [fill=gray, draw=black] (2.7,1.8) circle (0.7);
\filldraw [fill=gray, draw=black] (3.57,0.7) circle (0.7);
\filldraw [fill=gray, draw=black] (5.27,1.0) circle (1.0);
\end{tikzpicture}
&  
13~ & \begin{tikzpicture} [scale=0.2]
\draw (0.0,0.0) -- (6.27,0.0) -- (6.27,2.5) -- (0.0,2.5) -- (0,0);
\filldraw [fill=gray, draw=black] (1.0,1.0) circle (1.0);
\filldraw [fill=gray, draw=black] (2.7,0.7) circle (0.7);
\filldraw [fill=gray, draw=black] (3.57,1.8) circle (0.7);
\filldraw [fill=gray, draw=black] (5.27,1.5) circle (1.0);
\end{tikzpicture}
& ~$\tilde{V}_\mathrm{b}+2\tilde{V}_\mathrm{f}-3\tilde{V}_\mathrm{a}$
\\ \rule[-2mm]{0mm}{6mm}
14~ & \begin{tikzpicture} [scale=0.2]
\draw (0.0,0.0) -- (5.87,0.0) -- (5.87,2.5) -- (0.0,2.5) -- (0,0);
\filldraw [fill=gray, draw=black] (1.0,1.0) circle (1.0);
\filldraw [fill=gray, draw=black] (2.5,1.8) circle (0.7);
\filldraw [fill=gray, draw=black] (3.37,0.7) circle (0.7);
\filldraw [fill=gray, draw=black] (4.87,1.5) circle (1.0);
\end{tikzpicture}
& 
14~ & \begin{tikzpicture} [scale=0.2]
\draw (0.0,0.0) -- (5.87,0.0) -- (5.87,2.5) -- (0.0,2.5) -- (0,0);
\filldraw [fill=gray, draw=black] (1.0,1.5) circle (1.0);
\filldraw [fill=gray, draw=black] (2.5,0.7) circle (0.7);
\filldraw [fill=gray, draw=black] (3.37,1.8) circle (0.7);
\filldraw [fill=gray, draw=black] (4.87,1.0) circle (1.0);
\end{tikzpicture}
& ~$\tilde{V}_\mathrm{b}+2\tilde{V}_\mathrm{c}-3\tilde{V}_\mathrm{a}$
\\ \rule[-2mm]{0mm}{6mm}
15~ & \begin{tikzpicture} [scale=0.2]
\draw (0.0,0.0) -- (2.8,0.0) -- (2.8,2.5) -- (0.0,2.5) -- (0,0);
\filldraw [fill=gray, draw=black] (0.7,1.8) circle (0.7);
\filldraw [fill=gray, draw=black] (2.1,1.8) circle (0.7);
\end{tikzpicture}
&  
16~ & \begin{tikzpicture} [scale=0.2]
\draw (0.0,0.0) -- (2.8,0.0) -- (2.8,2.5) -- (0.0,2.5) -- (0,0);
\filldraw [fill=gray, draw=black] (0.7,0.7) circle (0.7);
\filldraw [fill=gray, draw=black] (2.1,0.7) circle (0.7);
\end{tikzpicture}
& ~$\tilde{V}_\mathrm{e}-\tilde{V}_\mathrm{a}$
\\ \rule[-2mm]{0mm}{6mm}
17~ & \begin{tikzpicture} [scale=0.2]
\draw (0.0,0.0) -- (3.14,0.0) -- (3.14,2.5) -- (0.0,2.5) -- (0,0);
\filldraw [fill=gray, draw=black] (0.7,0.7) circle (0.7);
\filldraw [fill=gray, draw=black] (1.57,1.8) circle (0.7);
\filldraw [fill=gray, draw=black] (2.44,0.7) circle (0.7);
\end{tikzpicture}
& 
17~ & \begin{tikzpicture} [scale=0.2]
\draw (0.0,0.0) -- (3.14,0.0) -- (3.14,2.5) -- (0.0,2.5) -- (0,0);
\filldraw [fill=gray, draw=black] (0.7,1.8) circle (0.7);
\filldraw [fill=gray, draw=black] (1.57,0.7) circle (0.7);
\filldraw [fill=gray, draw=black] (2.44,1.8) circle (0.7);
\end{tikzpicture}
&  ~$2\tilde{V}_\mathrm{b}-2\tilde{V}_\mathrm{a}$
\\ \hline\hline
\end{tabular}
\end{center}
\end{table}

\begin{table}[t]
  \caption{Species of quasiparticles which can be activated (directly or indirectly) from the reference state (\ref{eq:36}). The motifs shown are for $\sigma_\mathrm{L}=2$, $\sigma_\mathrm{S}=1.4$, $H=2.5$. The excess volumes $\Delta V_m$ are relative to a segment of reference state $\mathsf{pv_S}$ with the same number of disks. Species $m=7,8,13,14,17$ have two distinct motifs.}\label{tab:t7}
\begin{center}
\begin{tabular}{rl|rl|l} \hline\hline
$m$~ & motif & ~~$m$~ & motif & ~~$\Delta V_m$ \\ \hline \rule[-2mm]{0mm}{8mm}
1~ &\begin{tikzpicture} [scale=0.2]
\draw (0.0,0.0) -- (2.8,0.0) -- (2.8,2.5) -- (0.0,2.5) -- (0,0);
\filldraw [fill=gray, draw=black] (0.7,1.8) circle (0.7);
\filldraw [fill=gray, draw=black] (2.1,1.8) circle (0.7);
\end{tikzpicture}
 &  
2~ & \begin{tikzpicture} [scale=0.2]
\draw (0.0,0.0) -- (2.8,0.0) -- (2.8,2.5) -- (0.0,2.5) -- (0,0);
\filldraw [fill=gray, draw=black] (0.7,0.7) circle (0.7);
\filldraw [fill=gray, draw=black] (2.1,0.7) circle (0.7);
\end{tikzpicture}
& ~~$\tilde{V}_\mathrm{e}- \tilde{V}_\mathrm{b}$
\\ \rule[-2mm]{0mm}{6mm}
3~ & \begin{tikzpicture} [scale=0.2]
\draw (0.0,0.0) -- (4.4,0.0) -- (4.4,2.5) -- (0.0,2.5) -- (0,0);
\filldraw [fill=gray, draw=black] (0.7,1.8) circle (0.7);
\filldraw [fill=gray, draw=black] (2.2,1.0) circle (1.0);
\filldraw [fill=gray, draw=black] (3.7,1.8) circle (0.7);
\end{tikzpicture}
& 
4~ & \begin{tikzpicture} [scale=0.2]
\draw (0.0,0.0) -- (4.4,0.0) -- (4.4,2.5) -- (0.0,2.5) -- (0,0);
\filldraw [fill=gray, draw=black] (0.7,0.7) circle (0.7);
\filldraw [fill=gray, draw=black] (2.2,1.5) circle (1.0);
\filldraw [fill=gray, draw=black] (3.7,0.7) circle (0.7);
\end{tikzpicture}
 & ~~$2\tilde{V}_\mathrm{c}- 2\tilde{V}_\mathrm{b}$
\\ \rule[-2mm]{0mm}{6mm}
5~ & \begin{tikzpicture} [scale=0.2]
\draw (0.0,0.0) -- (4.75,0.0) -- (4.75,2.5) -- (0.0,2.5) -- (0,0);
\filldraw [fill=gray, draw=black] (0.7,1.8) circle (0.7);
\filldraw [fill=gray, draw=black] (2.37,1.5) circle (1.0);
\filldraw [fill=gray, draw=black] (4.05,1.8) circle (0.7);
\end{tikzpicture}
& 
6~ & \begin{tikzpicture} [scale=0.2]
\draw (0.0,0.0) -- (4.75,0.0) -- (4.75,2.5) -- (0.0,2.5) -- (0,0);
\filldraw [fill=gray, draw=black] (0.7,0.7) circle (0.7);
\filldraw [fill=gray, draw=black] (2.37,1.0) circle (1.0);
\filldraw [fill=gray, draw=black] (4.05,0.7) circle (0.7);
\end{tikzpicture}
& ~~$2\tilde{V}_\mathrm{f}- 2\tilde{V}_\mathrm{b}$
\\ \rule[-2mm]{0mm}{6mm}
7~ & \begin{tikzpicture} [scale=0.2]
\draw (0.0,0.0) -- (4.57,0.0) -- (4.57,2.5) -- (0.0,2.5) -- (0,0);
\filldraw [fill=gray, draw=black] (0.7,0.7) circle (0.7);
\filldraw [fill=gray, draw=black] (2.2,1.5) circle (1.0);
\filldraw [fill=gray, draw=black] (3.87,1.8) circle (0.7);
\end{tikzpicture}
 & 
7~ & \begin{tikzpicture} [scale=0.2]
\draw (0.0,0.0) -- (4.57,0.0) -- (4.57,2.5) -- (0.0,2.5) -- (0,0);
\filldraw [fill=gray, draw=black] (0.7,1.8) circle (0.7);
\filldraw [fill=gray, draw=black] (2.37,1.5) circle (1.0);
\filldraw [fill=gray, draw=black] (3.87,0.7) circle (0.7);
\end{tikzpicture}
& ~~$\tilde{V}_\mathrm{c}+V_\mathrm{f}- 2\tilde{V}_\mathrm{b}$
\\ \rule[-2mm]{0mm}{6mm}
8~ & \begin{tikzpicture} [scale=0.2]
\draw (0.0,0.0) -- (4.57,0.0) -- (4.57,2.5) -- (0.0,2.5) -- (0,0);
\filldraw [fill=gray, draw=black] (0.7,1.8) circle (0.7);
\filldraw [fill=gray, draw=black] (2.2,1.0) circle (1.0);
\filldraw [fill=gray, draw=black] (3.87,0.7) circle (0.7);
\end{tikzpicture}
 & 
8~ & \begin{tikzpicture} [scale=0.2]
\draw (0.0,0.0) -- (4.57,0.0) -- (4.57,2.5) -- (0.0,2.5) -- (0,0);
\filldraw [fill=gray, draw=black] (0.7,0.7) circle (0.7);
\filldraw [fill=gray, draw=black] (2.37,1.0) circle (1.0);
\filldraw [fill=gray, draw=black] (3.87,1.8) circle (0.7);
\end{tikzpicture}
 & ~~$\tilde{V}_\mathrm{c}+\tilde{V}_\mathrm{f}- 2\tilde{V}_\mathrm{b}$
\\ \rule[-2mm]{0mm}{6mm}
9~ & \begin{tikzpicture} [scale=0.2]
\draw (0.0,0.0) -- (6.54,0.0) -- (6.54,2.5) -- (0.0,2.5) -- (0,0);
\filldraw [fill=gray, draw=black] (0.7,1.8) circle (0.7);
\filldraw [fill=gray, draw=black] (2.37,1.5) circle (1.0);
\filldraw [fill=gray, draw=black] (4.34,1.0) circle (1.0);
\filldraw [fill=gray, draw=black] (5.84,1.8) circle (0.7);
\end{tikzpicture}
 & 
10~ & \begin{tikzpicture} [scale=0.2]
\draw (0.0,0.0) -- (6.54,0.0) -- (6.54,2.5) -- (0.0,2.5) -- (0,0);
\filldraw [fill=gray, draw=black] (0.7,0.7) circle (0.7);
\filldraw [fill=gray, draw=black] (2.2,1.5) circle (1.0);
\filldraw [fill=gray, draw=black] (4.16,1.0) circle (1.0);
\filldraw [fill=gray, draw=black] (5.84,0.7) circle (0.7);
\end{tikzpicture}
& ~~$\tilde{V}_\mathrm{a}+\tilde{V}_\mathrm{c}+\tilde{V}_\mathrm{f}- 3\tilde{V}_\mathrm{b}$
\\ \rule[-2mm]{0mm}{6mm}
11~ & \begin{tikzpicture} [scale=0.2]
\draw (0.0,0.0) -- (6.54,0.0) -- (6.54,2.5) -- (0.0,2.5) -- (0,0);
\filldraw [fill=gray, draw=black] (0.7,1.8) circle (0.7);
\filldraw [fill=gray, draw=black] (2.2,1.0) circle (1.0);
\filldraw [fill=gray, draw=black] (4.16,1.5) circle (1.0);
\filldraw [fill=gray, draw=black] (5.84,1.8) circle (0.7);
\end{tikzpicture}
 & 
12~ & \begin{tikzpicture} [scale=0.2]
\draw (0.0,0.0) -- (6.54,0.0) -- (6.54,2.5) -- (0.0,2.5) -- (0,0);
\filldraw [fill=gray, draw=black] (0.7,0.7) circle (0.7);
\filldraw [fill=gray, draw=black] (2.37,1.0) circle (1.0);
\filldraw [fill=gray, draw=black] (4.34,1.5) circle (1.0);
\filldraw [fill=gray, draw=black] (5.84,0.7) circle (0.7);
\end{tikzpicture}
& ~~$\tilde{V}_\mathrm{a}+\tilde{V}_\mathrm{c}+\tilde{V}_\mathrm{f}- 3\tilde{V}_\mathrm{b}$
\\ \rule[-2mm]{0mm}{6mm}
13~ & \begin{tikzpicture} [scale=0.2]
\draw (0.0,0.0) -- (6.71,0.0) -- (6.71,2.5) -- (0.0,2.5) -- (0,0);
\filldraw [fill=gray, draw=black] (0.7,1.8) circle (0.7);
\filldraw [fill=gray, draw=black] (2.37,1.5) circle (1.0);
\filldraw [fill=gray, draw=black] (4.34,1.0) circle (1.0);
\filldraw [fill=gray, draw=black] (6.01,0.7) circle (0.7);
\end{tikzpicture}
 & 
13~ & \begin{tikzpicture} [scale=0.2]
\draw (0.0,0.0) -- (6.71,0.0) -- (6.71,2.5) -- (0.0,2.5) -- (0,0);
\filldraw [fill=gray, draw=black] (0.7,0.7) circle (0.7);
\filldraw [fill=gray, draw=black] (2.37,1.0) circle (1.0);
\filldraw [fill=gray, draw=black] (4.34,1.5) circle (1.0);
\filldraw [fill=gray, draw=black] (6.01,1.8) circle (0.7);
\end{tikzpicture}
 & ~~$\tilde{V}_\mathrm{a}+2\tilde{V}_\mathrm{f}-3\tilde{V}_\mathrm{b}$
\\ \rule[-2mm]{0mm}{6mm}
14~ & \begin{tikzpicture} [scale=0.2]
\draw (0.0,0.0) -- (6.36,0.0) -- (6.36,2.5) -- (0.0,2.5) -- (0,0);
\filldraw [fill=gray, draw=black] (0.7,0.7) circle (0.7);
\filldraw [fill=gray, draw=black] (2.2,1.5) circle (1.0);
\filldraw [fill=gray, draw=black] (4.16,1.0) circle (1.0);
\filldraw [fill=gray, draw=black] (5.66,1.8) circle (0.7);
\end{tikzpicture}
 & 
14~ & \begin{tikzpicture} [scale=0.2]
\draw (0.0,0.0) -- (6.36,0.0) -- (6.36,2.5) -- (0.0,2.5) -- (0,0);
\filldraw [fill=gray, draw=black] (0.7,1.8) circle (0.7);
\filldraw [fill=gray, draw=black] (2.2,1.0) circle (1.0);
\filldraw [fill=gray, draw=black] (4.16,1.5) circle (1.0);
\filldraw [fill=gray, draw=black] (5.66,0.7) circle (0.7);
\end{tikzpicture}
 & ~~$\tilde{V}_\mathrm{a}+2\tilde{V}_\mathrm{c}- 3\tilde{V}_\mathrm{b}$
\\ \rule[-2mm]{0mm}{6mm}
15~ & \begin{tikzpicture} [scale=0.2]
\draw (0.0,0.0) -- (4.0,0.0) -- (4.0,2.5) -- (0.0,2.5) -- (0,0);
\filldraw [fill=gray, draw=black] (1.0,1.5) circle (1.0);
\filldraw [fill=gray, draw=black] (3.0,1.5) circle (1.0);
\end{tikzpicture}
& 
16~ & \begin{tikzpicture} [scale=0.2]
\draw (0.0,0.0) -- (4.0,0.0) -- (4.0,2.5) -- (0.0,2.5) -- (0,0);
\filldraw [fill=gray, draw=black] (1.0,1.0) circle (1.0);
\filldraw [fill=gray, draw=black] (3.0,1.0) circle (1.0);
\end{tikzpicture}
 & ~~$\tilde{V}_\mathrm{d}- \tilde{V}_\mathrm{b}$
\\ \rule[-2mm]{0mm}{6mm}
17~ & \begin{tikzpicture} [scale=0.2]
\draw (0.0,0.0) -- (5.93,0.0) -- (5.93,2.5) -- (0.0,2.5) -- (0,0);
\filldraw [fill=gray, draw=black] (1.0,1.0) circle (1.0);
\filldraw [fill=gray, draw=black] (2.96,1.5) circle (1.0);
\filldraw [fill=gray, draw=black] (4.93,1.0) circle (1.0);
\end{tikzpicture}
 & 
17~ & \begin{tikzpicture} [scale=0.2]
\draw (0.0,0.0) -- (5.93,0.0) -- (5.93,2.5) -- (0.0,2.5) -- (0,0);
\filldraw [fill=gray, draw=black] (1.0,1.5) circle (1.0);
\filldraw [fill=gray, draw=black] (2.96,1.0) circle (1.0);
\filldraw [fill=gray, draw=black] (4.93,1.5) circle (1.0);
\end{tikzpicture}
& ~~$2\tilde{V}_\mathrm{a}- 2\tilde{V}_\mathrm{b}$
\\ \hline\hline
\end{tabular}
\end{center}
\end{table}

%
\section{Combinatorics}\label{sec:comb}
%
The particles identified in Tables~\ref{tab:t3} or \ref{tab:t7} are statistically interacting in the sense that activating one particle from any species $n$ $(\Delta N_n=1)$ affects the number $d_m$ of remaining slots for the activation of further particles from each species $m$.
This type of interaction can be accounted for by a generalized Pauli principle, $\Delta d_m=-\sum_mg_{mn}\Delta N_n$, introduced by Haldane \cite{Hald91a} in a different context.
We employ this principle here in integrated form \cite{copic,janac1},
\begin{equation}\label{eq:5}
  d_m =A_m-\sum_{n=1}^M g_{mn}(N_{n}-\delta_{mn}),
\end{equation}
with capacity constants $A_m$ and statistical interaction coefficients $g_{mn}$ as listed in Table~\ref{tab:t4}.
Given a population of $N_n$ particles from all species $n$ already present, Eq.~(\ref{eq:5}) states that there are $d_m$ distinct ways of placing a particle of species $m$.
The entries for $A_m$ and $g_{mn}$ in Table~\ref{tab:t4} come from Ref.~\cite{pichs}, a study of the $s=\frac{3}{2}$ Ising chain, whose spin configurations can be mapped onto configurations of jammed disks of two sizes.

The capacity $A_m$ for compacts and hosts in the pseudo-vacua (\ref{eq:2}) or (\ref{eq:36}) is proportional to the number $N$ of disks in the channel.
Both reference states have zero capacity for tags, which can only be activated inside hosts.
Positive (negative) $g_{mn}$ produce negative (positive) changes $\Delta d_m$ in the number of open slots for particles $m$. 
Activating a compact $(n=1,2)$ reduces the number of open slots for the activation of hosts or further compacts ($m=1,\ldots,14$), but leaves the number of open slots for tags invariant.

The activation of a host has a similar effect on open slots for compacts or further hosts, except that open slots of some compacts or hosts are affected only indirectly.
Open slots for some tags are added when a host is activated.
Tags $m=15,16$ reduce the number of open slots for compacts or hosts, but leave the number of open slots for any tag invariant. 
Tag $m=17$, by contrast, increases the number of open slots for tags $m=15,16$.

The reference state (\ref{eq:2}) contains only large disks and the activation of particles from species $m=1,2$ keeps it that way. 
Particles from all other species replace one or two large disks by small disks.
This substitution is encoded in the quantum number $s_m$ of Table~\ref{tab:t4} for later use in the configurational statistical analysis.
When we start from reference state (\ref{eq:36}), the roles of large and small disks are interchanged.
The combinatorics remain the same.
The multiplicity of jammed microstates with particle content $\{N_m\}$ is a product of binomials \cite{Hald91a, copic, pichs}:
\begin{equation}\label{eq:6} 
W(\{N_m\})=n_\mathrm{pv}\prod_{m=1}^M\left(\begin{array}{c}d_m+N_m-1 \\ N_m\end{array}\right), 
\end{equation}
where the prefactor for open boundary conditions, $n_\mathrm{pv}=2$, has no bearing on the configurational statistics of macrosystems.
\begin{widetext}

\begin{table}[t]
  \caption{Specifications of $M=17$ species of particles with motifs compiled in Table~\ref{tab:t3}. These particles generate all jammed microstates from the reference state (\ref{eq:2}) or (\ref{eq:36}). The specifications are (from left to right) the species number $m$, the capacity constant $A_m$, the number $s_m$ of small or large disks added, and the interaction coefficients $g_{mn}$ between particles of all species. 
}\label{tab:t4} 
\begin{center}
{\begin{tabular}{r|ccc}
$m$  & $A_{m}$ & $s_m$
\\ \hline \rule[-2mm]{0mm}{6mm}
$1$ & $\frac{N-1}{2}$ & $0$ 
\\ \rule[-2mm]{0mm}{5mm}
$2$ & $\frac{N-1}{2}$ & $0$ 
\\ \rule[-2mm]{0mm}{5mm}
$3$ & $\frac{N}{2}-1$ & $1$ 
\\ \rule[-2mm]{0mm}{5mm}
$4$ & $\frac{N}{2}-1$ & $1$ 
\\ \rule[-2mm]{0mm}{5mm}
$5$ & $\frac{N}{2}-1$ & $1$ 
\\ \rule[-2mm]{0mm}{5mm}
$6$ & $\frac{N}{2}-1$ & $1$ 
\\ \rule[-2mm]{0mm}{5mm}
$7$ & $N-2$ & $1$ 
\\ \rule[-2mm]{0mm}{5mm}
$8$ & $N-2$ & $1$ 
\\ \rule[-2mm]{0mm}{5mm}
$9$ & $\frac{N-3}{2}$ & $2$
\\ \rule[-2mm]{0mm}{5mm}
$10$ & $\frac{N-3}{2}$ & $2$ 
\\ \rule[-2mm]{0mm}{5mm}
$11$ & $\frac{N-3}{2}$ & $2$ 
\\ \rule[-2mm]{0mm}{5mm}
$12$ & $\frac{N-3}{2}$ & $2$ 
\\ \rule[-2mm]{0mm}{5mm}
$13$ & $N-3$ & $2$ 
\\ \rule[-2mm]{0mm}{5mm}
$14$ & $N-3$ & $2$ 
\\ \rule[-2mm]{0mm}{5mm}
$15$ & $0$ & $1$ 
\\ \rule[-2mm]{0mm}{5mm}
$16$ & $0$ & $1$ 
\\ \rule[-2mm]{0mm}{5mm}
$17$ & $0$ & $2$ 
\end{tabular}\hspace{9mm}%
\begin{tabular}{r|rrrrrrrrrrrrrrrrr} 
$g_{mn}$ & $~1$ & $~2$ & $3$ & $4$ & $5$ & $6$ & $7$ & $8$ & $9$ & $10$ & 
$11$ & $12$ & $13$ & $14$ & $~15$ & $~16$ & $17$ \\
\hline \rule[-2mm]{0mm}{6mm}
$1$ & $\frac{1}{2}$ & $\frac{1}{2}$ & $0$ & $1$ 
& $0$ & $1$ & $\frac{1}{2}$ & $\frac{1}{2}$ 
& $\frac{1}{2}$ & $\frac{3}{2}$ & $\frac{1}{2}$ & $\frac{3}{2}$ 
& $1$ & $1$ & $\frac{1}{2}$ & $\frac{1}{2}$ 
& $1$ \\ \rule[-2mm]{0mm}{5mm}
$2$ & $\frac{1}{2}$ & $\frac{1}{2}$ & $1$ & $0$ 
& $1$ & $0$ & $\frac{1}{2}$ & $\frac{1}{2}$ 
& $\frac{3}{2}$ & $\frac{1}{2}$ & $\frac{3}{2}$ & $\frac{1}{2}$ 
& $1$ & $1$ & $\frac{1}{2}$ & $\frac{1}{2}$ 
& $1$ \\ \rule[-2mm]{0mm}{5mm}
$3$ & $\frac{1}{2}$ & $\frac{1}{2}$ & $1$ & $1$ 
& $1$ & $1$ & $\frac{1}{2}$ & $\frac{1}{2}$ 
& $\frac{1}{2}$ & $\frac{3}{2}$ & $\frac{1}{2}$ & $\frac{3}{2}$ 
& $1$ & $1$ & $\frac{1}{2}$ & $\frac{1}{2}$ 
& $1$ \\ \rule[-2mm]{0mm}{5mm}
$4$ & $\frac{1}{2}$ & $\frac{1}{2}$ & $1$ & $1$ 
& $1$ & $1$ & $\frac{1}{2}$ & $\frac{1}{2}$ 
& $\frac{3}{2}$ & $\frac{1}{2}$ & $\frac{3}{2}$ & $\frac{1}{2}$ 
& $1$ & $1$ & $\frac{1}{2}$ & $\frac{1}{2}$ 
& $1$ \\ \rule[-2mm]{0mm}{5mm}
$5$ & $\frac{1}{2}$ & $\frac{1}{2}$ & $0$ & $1$ 
& $1$ & $1$ & $\frac{1}{2}$ & $\frac{1}{2}$ 
& $\frac{1}{2}$ & $\frac{3}{2}$ & $\frac{1}{2}$ & $\frac{3}{2}$ 
& $1$ & $1$ & $\frac{1}{2}$ & $\frac{1}{2}$ 
& $1$ \\ \rule[-2mm]{0mm}{5mm}
$6$ & $\frac{1}{2}$ & $\frac{1}{2}$ & $1$ & $0$ 
& $1$ & $1$ & $\frac{1}{2}$ & $\frac{1}{2}$ 
& $\frac{3}{2}$ & $\frac{1}{2}$ & $\frac{3}{2}$ & $\frac{1}{2}$ 
& $1$ & $1$ & $\frac{1}{2}$ & $\frac{1}{2}$ 
& $1$ \\ \rule[-2mm]{0mm}{5mm}
$7$ & $1$ & $1$ & $2$ & $2$ 
& $2$ & $2$ & $2$ & $1$ 
& $3$ & $3$ & $3$ & $3$ 
& $3$ & $3$ & $1$ & $1$ 
& $2$ \\ \rule[-2mm]{0mm}{5mm}
$8$ & $1$ & $1$ & $2$ & $2$ 
& $2$ & $2$ & $2$ & $2$ 
& $3$ & $3$ & $3$ & $3$ 
& $3$ & $3$ & $1$ & $1$ 
& $2$ \\ \rule[-2mm]{0mm}{5mm}
$9$ & $\frac{1}{2}$ & $\frac{1}{2}$ & $1$ & $1$ 
& $1$ & $1$ & $\frac{1}{2}$ & $\frac{1}{2}$ 
& $\frac{3}{2}$ & $\frac{3}{2}$ & $\frac{1}{2}$ & $\frac{3}{2}$ 
& $1$ & $1$ & $\frac{1}{2}$ & $\frac{1}{2}$ 
& $1$ \\ \rule[-2mm]{0mm}{5mm}
$10$ & $\frac{1}{2}$ & $\frac{1}{2}$ & $1$ & $1$ 
& $1$ & $1$ & $\frac{1}{2}$ & $\frac{1}{2}$ 
& $\frac{3}{2}$ & $\frac{3}{2}$ & $\frac{3}{2}$ & $\frac{1}{2}$ 
& $1$ & $1$ & $\frac{1}{2}$ & $\frac{1}{2}$ 
& $1$ \\ \rule[-2mm]{0mm}{5mm}
$11$ & $\frac{1}{2}$ & $\frac{1}{2}$ & $1$ & $1$ 
& $1$ & $1$ & $\frac{1}{2}$ & $\frac{1}{2}$ 
& $\frac{3}{2}$ & $\frac{3}{2}$ & $\frac{3}{2}$ & $\frac{3}{2}$ 
& $1$ & $1$ & $\frac{1}{2}$ & $\frac{1}{2}$ 
& $1$ \\ \rule[-2mm]{0mm}{5mm}
$12$ & $\frac{1}{2}$ & $\frac{1}{2}$ & $1$ & $1$ 
& $1$ & $1$ & $\frac{1}{2}$ & $\frac{1}{2}$ 
& $\frac{3}{2}$ & $\frac{3}{2}$ & $\frac{3}{2}$ & $\frac{3}{2}$ 
& $1$ & $1$ & $\frac{1}{2}$ & $\frac{1}{2}$ 
& $1$ \\ \rule[-2mm]{0mm}{5mm}
$13$ & $1$ & $1$ & $2$ & $2$ 
& $2$ & $2$ & $1$ & $1$ 
& $3$ & $3$ & $3$ & $3$ 
& $3$ & $2$ & $1$ & $1$ 
& $2$ \\ \rule[-2mm]{0mm}{5mm}
$14$ & $1$ & $1$ & $2$ & $2$ 
& $2$ & $2$ & $1$ & $1$ 
& $3$ & $3$ & $3$ & $3$ 
& $3$ & $3$ & $1$ & $1$ 
& $2$ \\ \rule[-2mm]{0mm}{5mm}
$15$ & $0$ & $0$ & $0$ & $-1$ 
& $-1$ & $0$ & $-1$ & $0$ 
& $-1$ & $-1$ & $-1$ & $-1$ 
& $-1$ & $-1$ & $0$ & $0$ 
& $-1$ \\ \rule[-2mm]{0mm}{5mm}
$16$ & $0$ & $0$ & $-1$ & $0$ 
& $0$ & $-1$ & $0$ & $-1$ 
& $-1$ & $-1$ & $-1$ & $-1$ 
& $-1$ & $-1$ & $0$ & $0$ 
& $-1$ \\ \rule[-2mm]{0mm}{5mm}
$17$ & $0$ & $0$ & $-1$ & $-1$ 
& $-1$ & $-1$ & $-1$ & $-1$ 
& $-1$ & $-1$ & $-1$ & $-1$ 
& $-1$ & $-1$ & $0$ & $0$ 
& $0$ \\ 
\end{tabular}}
\end{center}
\end{table} 

\end{widetext}

%
\section{Energetics}\label{sec:ener}
%
All ingredients to our methodology discussed thus far pertain to jammed microstates: configurations of statistically interacting particles activated from reference states (\ref{eq:2}) or (\ref{eq:36}), countable by the multiplicity expression (\ref{eq:6}).
The jamming protocol, specifically the relevant energies that govern the disks prior to jamming, determines the statistical weight of jammed microstates in the analysis of configurational statistics.

In this study, we consider two kinds of energy, both present prior to jamming:
\begin{itemize}

\item[--] The kinetic energy associated with random agitations.
This form of energy has a controllable intensity $T_\mathrm{k}$.
It is the granular-system equivalent of the thermal energy $k_\mathrm{B}T$, such as applicable to colloids.

\item[--] The potential energy associated with work against the pistons at the ends of the channel. 
This form of energy is encoded in a set of distinct activation energies $\epsilon_m$ for each particle species as will be discussed later (Sec.~\ref{sec:jam-prot}) in some detail.

\end{itemize}

The $\epsilon_m$ at given $T_\mathrm{k}$ govern the propensity of specific disk configurations to be realized in jammed microstates.

The composition of jammed macrostates, including their volume, entropy, and mix of disk sizes, is thus determined by geometry, combinatorics, and energetics.
Multiple scenarios are amenable to an analytic solution or a precision numerical solution.
All scenarios share the same geometry and combinatorics.
The energetics make them diverse by the choice of activation energies $\epsilon_m$ associated with particular jamming protocols. 
Each microstate is assigned an energy expression in the form of a sum of activation energies of all the particles it contains:
\begin{equation}\label{eq:4} 
E(\{N_m\})=E_\mathrm{pv}+\sum_{m=1}^MN_m\epsilon_m,
\end{equation}
where $E_\mathrm{pv}$ is the reference energy of state (\ref{eq:2}) or state (\ref{eq:36}).
Expressions (\ref{eq:6}) for multiplicity and (\ref{eq:4}) for energy with specifications from Tables~\ref{tab:t2}-\ref{tab:t4} are the main ingredients to the configurational statistical analysis.

%
\section{Configurational statistics}\label{sec:stat}
%
The mathematical structure of the analysis of statistically interacting particles adapted to the configurational statistics of jammed disks was developed in Refs.~\cite{Wu94, Isak94} for thermal systems and in Refs.~\cite{janac1, janac2, janac3} for configurational statistics.
Here we build on these advances for the analysis of macroscopic systems in the present context. 
The thermodynamic limit at the most fundamental level concerns the capacity constants and affects primarily the average particle content of jammed macrostates:
\begin{equation}\label{eq:13}
\bar{A}_m\doteq\lim_{N\to\infty}\frac{A_m}{N}, \quad 
\bar{N}_m\doteq\lim_{N\to\infty}\frac{\langle N_m\rangle}{N}.
\end{equation}
Two quantities of interest, expressed as functions of particle content, are the excess volume and the configurational entropy:
\begin{equation}\label{eq:7} 
\bar{V}\doteq\lim_{N\to\infty}\frac{V-V_\mathrm{pv}}{N}=\sum_m\bar{N}_m\Delta V_m,
\end{equation}
\begin{align}\label{eq:20}
\bar{S} = \lim_{N\to\infty}\frac{S}{k_B}
&=\sum_m\Big[\big(\bar{N}_{m}
+\bar{Y}_m\big)\ln\big(\bar{N}_m+\bar{Y}_m\big) \nonumber \\
 &\hspace{14mm} -\bar{N}_m \ln \bar{N}_m -\bar{Y}_m\ln \bar{Y}_m\Big],
 \nonumber \\ 
\bar{Y}_m &\doteq \bar{A}_m-\sum_mg_{mn} \bar{N}_{n}.
\end{align}
The mathematical analysis relies on the invertibility of the (generally asymmetric) matrix,
\begin{equation}\label{eq:14}
G_{mn}=g_{mn}+w_m\delta_{mn},
\end{equation}
constructed from the (known) interaction coefficients $g_{mn}$ and the (yet unknown) variables $w_{m}$.
It delivers the population densities of particles in jammed macrostates as the solution of a set of linear equations,
\begin{equation}\label{eq:15}
\sum_nG_{mn}\bar{N}_n=\bar{A}_m\quad \Rightarrow~
\bar{N}_n=\sum_mG^{-1}_{nm}\bar{A}_m.
\end{equation}
The $w_m$ are the physically relevant solutions of the set of nonlinear algebraic equations,
\begin{equation}\label{eq:16} 
e^{\beta\epsilon_m}=(1+w_m)\prod_n \big(1+w_{n}^{-1}\big)^{-g_{nm}},
\end{equation}
where we have introduced the parameter $\beta=T_\mathrm{k}^{-1}$ akin to an inverse scaled temperature in statistical mechanics, but here representing a measure for the intensity of random agitations prior to jamming.

By transforming the $w_m$ to new variables $x_m$ and introducing energy parameters $e_m$,
\begin{equation}\label{eq:17}
x_m\doteq\frac{w_m}{1+w_m},\quad e_m\doteq e^{-\beta\epsilon_m},
\end{equation}
these algebraic equations acquire the form of fixed-point equations,
\begin{equation}\label{eq:19}
x_m=1-e_m\prod_nx_n^{g_{nm}},
\end{equation}
with solutions $x_m\big(\{e_n\}\big)$ of range $0\leq x_m\leq 1$.
This rendering is advantageous in the numerical analysis.
The partition function $\bar{Z}$ or the thermodynamic potential $\bar{\Omega}$,
\begin{equation}\label{eq:18}
\bar{Z}=\prod_mx_m^{-\bar{A}_m},\quad \bar{\Omega}=-\beta^{-1}\ln\bar{Z},
\end{equation}
determine the physical quantities of interest here.
For population densities and entropy we write
\begin{equation}\label{eq:21}
\bar{N}_m=\frac{\partial\ln\bar{Z}}{\partial\ln e_m}
=-\sum_n\bar{A}_n\frac{\partial\ln x_n}{\partial\ln e_m},
\end{equation}
\begin{equation}\label{eq:22}
\bar{S}=\beta^2\frac{\partial\bar{\Omega}}{\partial\beta}
=-\sum_m\Big[\bar{N}_m\ln e_m+\bar{A}_m\ln x_m\Big]. 
\end{equation}

Experimentally, the channel would contain fixed numbers of large and small disks. 
This corresponds to a canonical ensemble. 
Our methodology operates in the grandcanonical ensemble, where average numbers of large and small disks are determined by the activation energies $\epsilon_m$.
We control the fractions $\bar{N}_\mathrm{S}$ and $\bar{N}_\mathrm{L}=1-\bar{N}_\mathrm{S}$ of small and large disks, respectively, by the amendment, $-\mu s_m$, to the activation energy $\epsilon_m$.
In this amendment, $\mu$ is a chemical potential of sorts and the quantum number $s_m$ (see Table~\ref{tab:t4}) counts the number of small disks in particles of species $m$ activated from reference state (\ref{eq:2}).

The partition function (\ref{eq:18}) with control variables $\beta$ and $\mu$ produces a unique functional relation, 
\begin{equation}\label{eq:8}
\bar{N}_\mathrm{S}(\beta,\mu)=\sum_ms_m\bar{N}_m,
\end{equation}
which allows us to keep $\bar{N}_\mathrm{S}$ fixed for any value of $\beta$, independent of the $\epsilon_m$.
An equivalent expression obtains for $\bar{N}_\mathrm{L}(\beta,\mu)$ if the reference state is (\ref{eq:36}).
All the results presented in Sec.~\ref{sec:scen1} pertain to specific fractions $\bar{N}_\mathrm{S}$ of small disks, understood as averages for a given yet not stated value of the control variable $\mu$. Recall that the thermodynamic limit is implied in (\ref{eq:13}), which ensures ensemble equivalence.

%
\section{Emphasis on symmetry}\label{sec:scen1}
%
We now apply this extended methodology of configurational statistical analysis to the scenario of maximum symmetry, which originates in the mapping to the $s=\frac{3}{2}$ Ising chain as explained in Sec.~\ref{sec:comb}.
Its one drawback (to be addressed in Sec.~\ref{sec:scen2}) is far outweighed by the mathematical simplicity and transparency. 
It can be analyzed in two versions, one starting from reference state (\ref{eq:2}) and the other from reference state (\ref{eq:36}).
Both versions produce identical results for volume and entropy.

\subsection{Reference state}\label{sec:scen1L}
We employ the large-disk reference state $\mathsf{pv_L}$ of Eq.~(\ref{eq:2}), from which particles with specifications compiled in Tables~\ref{tab:t3} and \ref{tab:t4} are activated.
We note that a particle with index  $m=1,2,15,16$ has only marginal stability if it immediately succeeds a particle of the same index and that particles $m=5,6$ are unstable in some configurations (see Tables~\ref{tab:t1} and \ref{tab:t3}). 

The first benefit of symmetry is that 6 pairs of particle species can each be merged.
The 11 remaining species (down from 17) have combinatorial specifications as compiled in Table~\ref{tab:t5}.  
The general conditions for such mergers were established by Anghel \cite{Anghel} and explained in Ref.~\cite{pichs} for Ising chains. 
All pairs of species that can be merged in this scenario have successive odd and even indices $m$, of which we keep the odd index after the merger: $m=1, 3, 5, 9, 11, 15$.
The motifs of merged species are not interchangeable.
Only one motif fits into a given open slot as is the case for (unmerged) species $m=7,8,13,14,17$ with two motifs each.

\begin{table}[t]
  \caption{Scaled capacity constants $\bar{A}_m$, quantum numbers $s_m$, and interaction coefficients $\hat{g}_{mn}$ of the particle species after merging operations. The $\Delta V_m$ are as in Table~\ref{tab:t3}.}\label{tab:t5}
\begin{center}
\begin{tabular}{r|rr}
$m$ & $\bar{A}_m$ & $s_m$\\ \hline
1 & 1 & 0 \\
3 & 1 & 1 \\
5 & 1 & 1 \\
7 & 1 & 1 \\
8 & 1 & 1 \\
9 & 1 & 2 \\
11 & 1 & 2 \\
13 & 1 & 2 \\
14 & 1 & 2 \\
15 & 0 & 1 \\
17 & 0 & 2 \\
\end{tabular} \hspace*{5mm}
\begin{tabular}{r|rrrrrrrrrrr} 
  $\hat{g}_{mn}$
    &~1 &~3 &~5 &~7 &~8 &~9 &11& 13& 14& 15& 17\\ \hline
  1 & 1 & 1 & 1 & 1 & 1 & 2 & 2 & 2 & 2 & 1 & 2\\ 
  3 & 1 & 2 & 2 & 1 & 1 & 2 & 2 & 2 & 2 & 1 & 2\\
  5 & 1 & 1 & 2 & 1 & 1 & 2 & 2 & 2 & 2 & 1 & 2\\
  7 & 1 & 2 & 2 & 2 & 1 & 3 & 3 & 3 & 3 & 1 & 2\\
  8 & 1 & 2 & 2 & 2 & 2 & 3 & 3 & 3 & 3 & 1 & 2\\
  9 & 1 & 2 & 2 & 1 & 1 & 3 & 2 & 2 & 2 & 1 & 2\\
  11& 1 & 2 & 2 & 1 & 1 & 3 & 3 & 2 & 2 & 1 & 2\\
  13& 1 & 2 & 2 & 1 & 1 & 3 & 3 & 3 & 2 & 1 & 2\\   
  14& 1 & 2 & 2 & 1 & 1 & 3 & 3 & 3 & 3 & 1 & 2\\
  15& 0 &-1 &-1 &-1 &-1 &-2 &-2 &-2 &-2 & 0 &-2\\   
  17& 0 &-1 &-1 &-1 &-1 &-1 &-1 &-1 &-1 & 0 & 0\\   
\end{tabular}
\end{center}
\end{table} 

We can express the activation energies of this scenario with no prejudice or bias in the form
\begin{equation}\label{eq:12}
\epsilon_m=p_m-\mu s_m.
\end{equation}
The variables $p_m$ represent work against the pressure of the pistons exerted prior to jamming.
The variable $\mu$ controls the fraction of small disks in the channel as explained earlier.
The $p_m$ are determined by the jamming protocol (Sec.~\ref{sec:jam-prot} below).

\subsection{Partition function}\label{sec:Z1L}
The second benefit of symmetry is a relation between the partition function $\bar{Z}$ in the product form (\ref{eq:18}) and the solution $x_1$ of Eqs.~(\ref{eq:19}) with energy parameter $e_1$ as defined in (\ref{eq:17}): 
\begin{equation}\label{eq:40}
\bar{Z}=\frac{e_1}{1-x_1}.
\end{equation}
This simplification (worked out in Appendix~\ref{sec:appa}) is due to the equality, $\bar{A}_m=\hat{g}_{m1}$, in Table~\ref{tab:t5}.

The third benefit of symmetry is also associated with the structure of the matrix $\hat{g}_{mn}$.
We can solve Eqs.~(\ref{eq:19}) iteratively for all $x_m$ with $m\neq1$ in terms of $x_1$ if we proceed in the sequence of indices $m=15, 17, 8, 7, 5, 3, 9, 11, 13, 14$.
The result is a polynomial equation for $\zeta\doteq\bar{Z}^{-1}$.
This polynomial is only of third order -- a fourth benefit of symmetry.
A unique physical solution can be identified under mild constraints on the energy parameters $e_m$ from (\ref{eq:17}), here modified by the fugacity $z$ as follows:
\begin{equation}\label{eq:25}
\bar{e}_m=e^{-\beta p_m}=e_mz^{-s_m},\quad z\doteq e^{\beta\mu}.
\end{equation}
From the expression for the partition thus calculated,
  \begin{align}\label{eq:26}
    \bar{Z} &= \frac{1}{2}\Big[
    1+\bar{e}_{1}+(\bar{e}_{15}+\sqrt{\bar{e}_{17}})z \\ \nonumber  
    &+
    \sqrt{\big(1+\bar{e}_{1}-(\bar{e}_{15}+\sqrt{\bar{e}_{17}})z\big)^{2}
      -4z(\bar{e}_{3}+2\bar{e}_{8}+\bar{e}_{5})}\,\Big],
  \end{align}
we can infer, by way of the relations in Sec.~\ref{sec:stat}, explicit results for quantities of interest in systems with a fixed fraction of small disks.

\subsection{Energy parameters}\label{sec:jam-prot}
The analytic solution is exact within the framework of configurational statistics.
It can be completed on that level of rigor for arbitrary activation energies (\ref{eq:12}).
However, physically meaningful results require specific values for the contributions $p_m$.
These values encode the pressure work of the disks against the pistons in the state of random agitations of given intensity $T_\mathrm{k}=\beta^{-1}$, before jamming is precipitated by a much higher pressure.
It is in this part of the modeling where additional assumptions come into play.

In previous applications of the same methodology to disks jammed in narrow channels \cite{janac1, janac2, janac3}, highly plausible values could be provided with confidence because the jamming protocols involved no variation in the channel width.
In the applications to be tackled here, we must allow small and large disks to pass each other during the phase of agitations, which complicates and diversifies possible jamming protocols.

Shuffling disks of two sizes by random agitations requires the channel width prior to jamming to exceed the value ${\sigma_\mathrm{L}+\sigma_\mathrm{S}}$ at the very least.
The jamming protocol thus combines two components (A) and (B), which can be implemented in multiple ways.

\begin{itemize}

\item[--]Component (A) increases the pressure from a moderate value $P$ to a much higher value that stops all motion.

\item[--] Component (B) narrows the channel from a width that allows the exchange of disk positions in the sequence down to the jamming width $H$.

\end{itemize}

In combination, (A) and (B) freeze out a jammed configuration of large and small disks.
The probability distribution of such jammed configurations depends on the two-component jamming protocol via energy parameters (\ref{eq:25}) with values that are not easy to pin down.

The alternative of leaving the channel width in excess of the value $\sigma_\mathrm{L}+\sigma_\mathrm{S}$ upon jamming would simplify the protocol, but at the high cost of making the diversity of jammed configurations unmanageable for exact analysis.  

In Refs.~\cite{janac1,janac2, janac3}, jamming only involved component (A).
The natural choice for the first term in Eq.~(\ref{eq:12}) was,
\begin{equation}\label{eq:39}
p_m=P\Delta V_m,
\end{equation}
where $\Delta V_m$ is an attribute of the jammed state, namely the excess volume of particles from species $m$ relative to a segment of reference state.
This choice was justified by the fact that the disks have very little wiggle room away from jammed configurations, just enough for each disk to move from one wall to the other.

For this work we proceed under two main assumptions: (i) the $p_m$ in the form (\ref{eq:39}) with the $\Delta V_m$ from Table~\ref{tab:t3} are still justifiable for a jamming protocol in which component (B) finishes before component (A); (ii) the effects of different jamming protocols on the selection of jammed configurations can be accounted for by variations in the parameters $\Delta V_m$ from their default values given in Table~\ref{tab:t3}.

These default values serve as a convenient reference point for the energy parameters that determine the jammed macrostate. 
Given that we can associate them with a particular type of jamming protocol, it is most significant that it takes only small variations among the $\Delta V_m$ (in the 5\% range) to produce different ordering tendencies, as further explained below.

Irrespective of how the $\Delta V_m$ are being selected, it is convenient to set $P=1$, which converts excess volumes into measures of expansion work against the piston prior to jamming.

\subsection{Steric ordering tendencies}\label{sec:ssm}
An exploration of the parameter space around the default values for the $\Delta V_m$ stated in Table~\ref{tab:t3} reveals that there are two regimes with distinct ordering tendencies, \emph{size segregation} and \emph{size alternation}, both associated with low entropy.
On the border between the two regimes, a tendency toward \emph{size randomness}, associated with high entropy, prevails.

Connecting jamming protocols with a specific ordering tendency is a difficult task, requiring a major effort in stochastic modeling or computer simulation, which is beyond the scope of this work.
Our methodology merely connects hypothetical jamming protocols with probabilities of jammed macrostates via energy parameters that our encoded in the $\Delta V_m$.

The switch between the two regimes is triggered by a 1-parameter variation in the $\Delta V_m$.
This variation only involves configurations of two large disks and two small disks which, upon jamming, become
\begin{equation}\label{eq:30}
\begin{tikzpicture} 
    \definecolor{mygray}{RGB}{100,100,100}
    \colorlet{lightblue}{blue!60}
    
    \def\scalefactor{0.2} 
    \def\larrowlength{0.4} 
    \def\rarrowlength{0.4} 
    \def\arrowlen{0.7cm} 
    \def\laboffset{0.2cm} 
    \def\boxheightscaled{2.5 * \scalefactor} 
    \def\verticalspacing{0.01cm} 
    \def\deltanuLineVext{0.01cm} 
    \def\deltanuYoffsetBelowBoxes{0.008cm} 

    \def\boxwidthtop{6.2 * \scalefactor} 
    \def\boxwidthbottom{6.01 * \scalefactor} 

    \begin{scope}[shift={(0,0)}, scale=\scalefactor]
        \draw (0,0) -- (0,2.5);
        \draw (0,0) -- (6.01,0);
        \draw (0,2.5) -- (6.01,2.5);
        \draw (6.01,0) -- (6.01,2.5);
        \filldraw [fill=gray, draw=black] (1.0,1.5) circle (1.0);
        \filldraw [fill=gray, draw=black] (2.94,1.0) circle (1.0);
        \filldraw [fill=gray, draw=black] (4.44,1.8) circle (0.7);
        \filldraw [fill=gray, draw=black] (5.31,0.7) circle (0.7);
    \end{scope}
    \coordinate (globalBottomBoxMidRight) at (\boxwidthbottom, 0.5 * \boxheightscaled);

    \draw[-] (globalBottomBoxMidRight) -- ++(0,0) node[right=\laboffset] (labelBottomText) {$\mathbf{\tilde{V}_{\mathrm{a}} + \tilde{V}_{\mathrm{b}}}$};
    
    \draw[-] (globalBottomBoxMidRight) -- ++(0,-0.6) node[right=\laboffset] (labelBottomText) {$\mathbf{\Delta \mathcal{V}}$};

    \pgfmathsetmacro{\topBoxGlobalShiftY}{\boxheightscaled + \verticalspacing}
    \begin{scope}[shift={(0, \topBoxGlobalShiftY)}, scale=\scalefactor]
        \draw (0,0) -- (0,2.5);
        \draw (0,0) -- (6.2,0);
        \draw (0,2.5) -- (6.2,2.5);
        \draw (6.2,0) -- (6.2,2.5);
        \filldraw [fill=gray, draw=black] (1,1.5) circle (1.0);
        \filldraw [fill=gray, draw=black] (2.5,0.7) circle (0.7);
        \filldraw [fill=gray, draw=black] (4.0,1.5) circle (1.0);
        \filldraw [fill=gray, draw=black] (5.5,0.7) circle (0.7);
    \end{scope}
    \coordinate (globalTopBoxMidRight) at (\boxwidthtop, \topBoxGlobalShiftY + 0.5 * \boxheightscaled);

    \draw[-] (globalTopBoxMidRight) -- ++(0,0) node[right=\laboffset] (labelTopText) {$\mathbf{2\tilde{V}_{\mathrm{c}}}$};


    \pgfmathsetmacro{\deltanuLineTotalYStart}{-\deltanuYoffsetBelowBoxes - \deltanuLineVext} 
    \pgfmathsetmacro{\deltanuLineTotalYEnd}{\topBoxGlobalShiftY + \boxheightscaled + \deltanuLineVext} 

    \draw[lightblue, thin] (\boxwidthbottom, \deltanuLineTotalYStart) -- (\boxwidthbottom, \deltanuLineTotalYEnd);
    \draw[lightblue, thin] (\boxwidthtop, \deltanuLineTotalYStart) -- (\boxwidthtop, \deltanuLineTotalYEnd);
    
    \pgfmathsetmacro{\deltanuCenterY}{-\deltanuYoffsetBelowBoxes}

\end{tikzpicture}
\end{equation}
These excess volumes satisfy the inequality,
\begin{equation}\label{eq:31}
 \Delta\mathcal{V}\doteq 2\tilde{V}_\mathrm{c}-(\tilde{V}_\mathrm{a}+\tilde{V}_\mathrm{b})>0.
\end{equation}
which, as it turns out, favors size segregation.
However, different jamming protocols are likely associated with energy parameters for which $\Delta\mathcal{V}<0$ is the best match, in which case the ordering tendency switches to size alternation.

The two ordering tendencies (segregation or alternation) are dominated by tiles 1 to 8 from Table~\ref{tab:t1}. 
These tiles form domains of the most compact configurations for given $0<\bar{N}_\mathrm{S}<1$.
Domains of size-segregated disks are sequences of tiles 1 to 4 and domains in a size-alternating pattern are sequences of tiles 5 to 8. 
The role of tiles 9 to 16 is reduced to mark walls between such domains. 

If the jamming protocol calls for $\Delta\mathcal{V}>0$, the energetically favored configuration of two large and two small disks is the one which, upon jamming is shown at the bottom in (\ref{eq:30}).
The jammed state frozen out of the lowest-energy unjammed state has all disks size-segregated.
This type of ordering has zero entropy upon completion irrespective of the fraction of small disks.

On the other hand, if the jamming protocol calls for $\Delta\mathcal{V}<0$, the energetically favored configuration of two small and two large disks is the top one in (\ref{eq:30}).
The jammed state frozen out of the lowest-energy unjammed state will exhibit an alternating pattern of a length determined by the fraction of small or large disks, whichever are in the minority.
The unpaired majority-size disks provide a medium for the ordered pairs spread out in a random arrangement and produce a residual entropy.

In the following, we work out three representative cases of jamming protocols, the first demanding $\Delta\mathcal{V}>0$, which favors size segregation, the second demanding $\Delta\mathcal{V}<0$, which favors size alternation, and the third positioned on the border $(\Delta\mathcal{V}=0)$, which favors neither ordering tendency.

\subsection{Size segregation}\label{sec:ssm-a}
For the first case we use energy parameters inferred from the default $\Delta V_m$-values listed in Table~\ref{tab:t3}, which imply $\Delta\mathcal{V}>0$ according to (\ref{eq:31}).
With this choice, the analysis that lead to the result (\ref{eq:26}) for the partition function can be continued.
It is worked out in Appendix~\ref{sec:appa}. 
Explicit results for population densities, excess volume, and entropy are presented and interpreted in the following.

\begin{figure}[t]
  \begin{center}
\includegraphics[width=43mm]{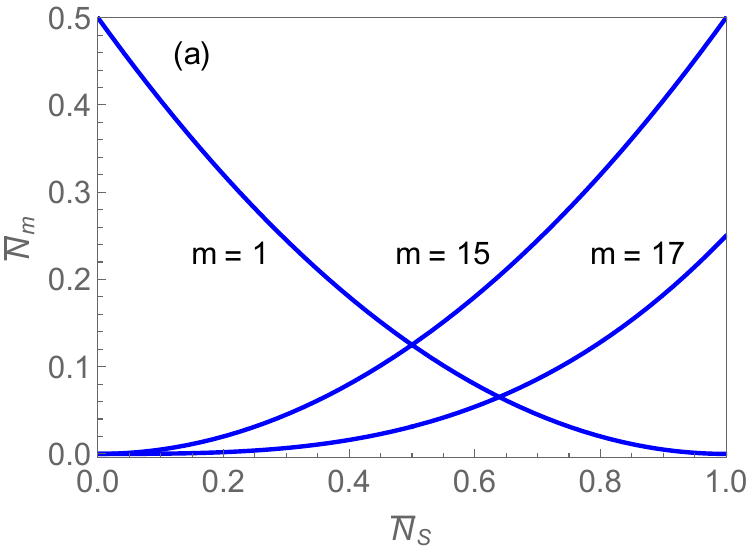}%
\hspace*{1mm}\includegraphics[width=43mm]{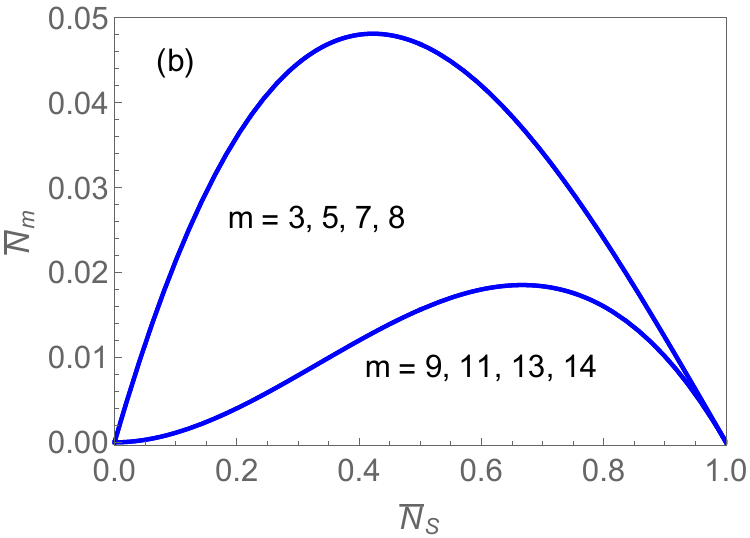}
\end{center}
\caption{Population densities (\ref{eq:27}) of particles from species $m=1,3,\ldots,17$ versus the fraction $\bar{N}_\mathrm{S}$ of small disks in the limit $\beta=0$. Note the different vertical scales in (a) and (b).}
  \label{fig:A1}
\end{figure}

\begin{figure}[b]
  \begin{center}
\includegraphics[width=43mm]{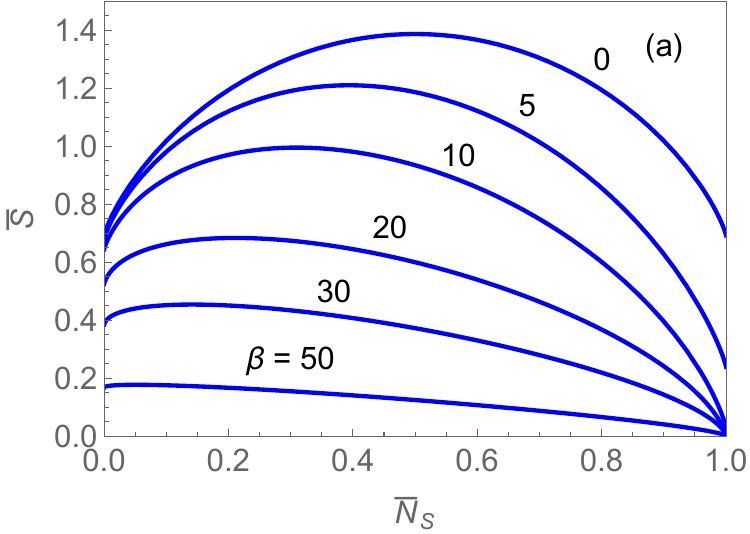}%
\hspace*{1mm}\includegraphics[width=43mm]{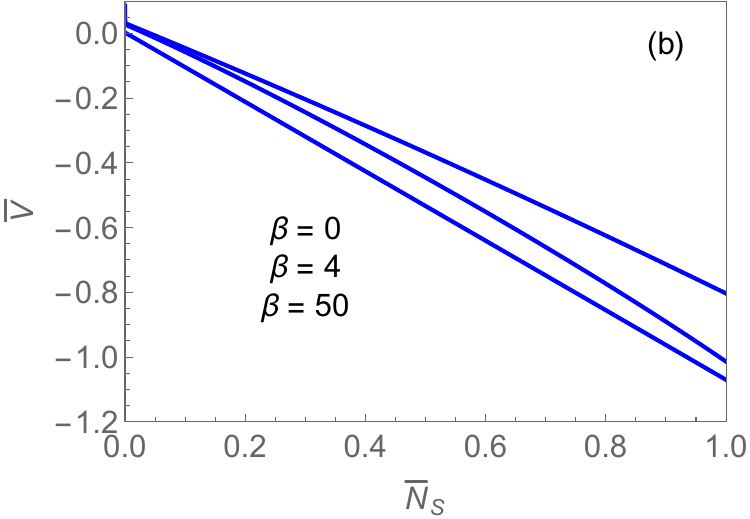}
\end{center}
\caption{(a) Entropy $\bar{S}$ and (b) excess volume $\bar{V}$ versus the fraction $\bar{N}_\mathrm{S}$ of small disks for various values of $\beta$.}
  \label{fig:A2}
\end{figure}

The limit $\beta=0$ pertains to high-intensity agitations prior to jamming.
Here the energy parameters play no part. 
The jammed macrostate only depends on the fraction $\bar{N}_\mathrm{S}$ of small disks in the channel.
The population densities for particles from each species are
\begin{align}\label{eq:27}
& \bar{N}_1=\frac{1}{2}(1-\bar{N}_\mathrm{S})^2, 
\nonumber \\
&\bar{N}_3=\bar{N}_5=\bar{N}_7=\bar{N}_8
=\frac{1}{8}\bar{N}_\mathrm{S}(2-\bar{N}_\mathrm{S})(1-\bar{N}_\mathrm{S}), 
\nonumber \\
&\bar{N}_9=\bar{N}_{11}=\bar{N}_{13}=\bar{N}_{14}
=\frac{1}{8}\bar{N}_\mathrm{S}^2(1-\bar{N}_\mathrm{S}),
\nonumber \\
&\bar{N}_{15}=\frac{1}{2}\bar{N}_\mathrm{S}^2,\quad
\bar{N}_{17}=\frac{1}{4}\bar{N}_\mathrm{S}^3.
\end{align}
As shown in Fig.~\ref{fig:A1}, particles 1 are dominant when most disks are large, whereas particles 15, 17 dominate when most disks are small.
The remaining species are activated in smaller numbers if disks of both sizes are present.
The excess volume (relative to the state $\mathsf{pv_L}$) and the entropy follow directly from these $\bar{N}_m$ via (\ref{eq:7}) and (\ref{eq:20}). 
The dependence on $\bar{N}_\mathrm{S}$ can be stated compactly:
\begin{subequations}

\begin{align}\label{eq:28}
& \bar{V}=v_0-v_1\bar{N}_\mathrm{S}-v_2\bar{N}_\mathrm{S}^2, \\
& v_0\doteq\frac{1}{2}(\tilde{V}_\mathrm{d}-\tilde{V}_\mathrm{a}),\quad 
v_1=\tilde{V}_\mathrm{a}+\tilde{V}_\mathrm{d}-\tilde{V}_\mathrm{c}-\tilde{V}_\mathrm{f}, \\
& v_2=\tilde{V}_\mathrm{c}+\tilde{V}_\mathrm{f}-\frac{1}{2}(\tilde{V}_\mathrm{a}
+\tilde{V}_\mathrm{b}+\tilde{V}_\mathrm{d}+\tilde{V}_\mathrm{e}),
\end{align}
\end{subequations}
\begin{equation}\label{eq:29}
\bar{S}=\ln2-\bar{N}_\mathrm{S}\ln\bar{N}_\mathrm{S}
-(1-\bar{N}_\mathrm{S})\ln(1-\bar{N}_\mathrm{S}).
\end{equation}
The top curve in each panel of Fig.~\ref{fig:A2} shows these results.
The entropy $\bar{S}$ is mirror-symmetric about $\bar{N}_\mathrm{S}=\frac{1}{2}$.
The excess volume $\bar{V}$ shrinks monotonically with increasing $\bar{N}_\mathrm{S}$ from $v_0$ as small disks replace large disks.

Reducing the intensity of agitations prior to jamming has a stronger effect on the entropy than on the excess volume. The trends are shown in Fig.~\ref{fig:A2}.
The rate at which the excess volume shrinks increases somewhat as the more compact disk configurations are realized with higher probability.
The growing preponderance of compact disk configurations reduces the entropy for all values of $\bar{N}_\mathrm{S}$.
The mirror symmetry is broken.
Small and large disks affect the energy parameters of particles differently.
The entropy approaches zero across the range of $\bar{N}_\mathrm{S}$, an unmistakable sign of ordering.

\begin{figure}[t]
  \begin{center}
\includegraphics[width=43mm]{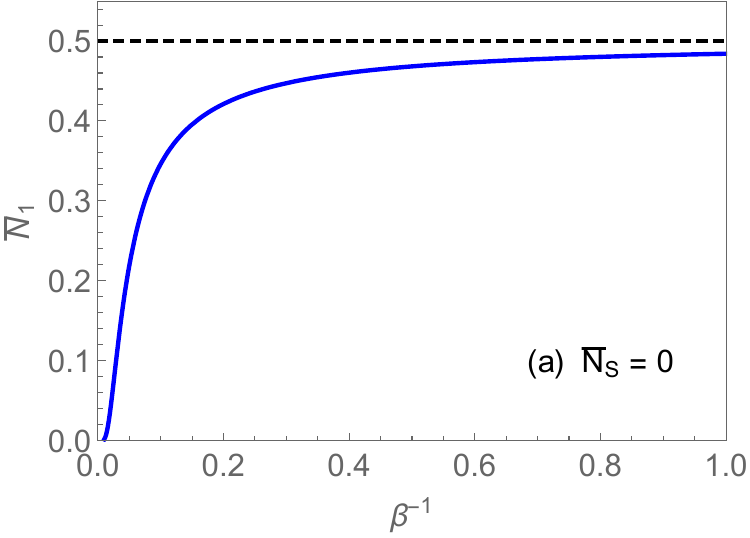}%
\hspace*{1mm}\includegraphics[width=43mm]{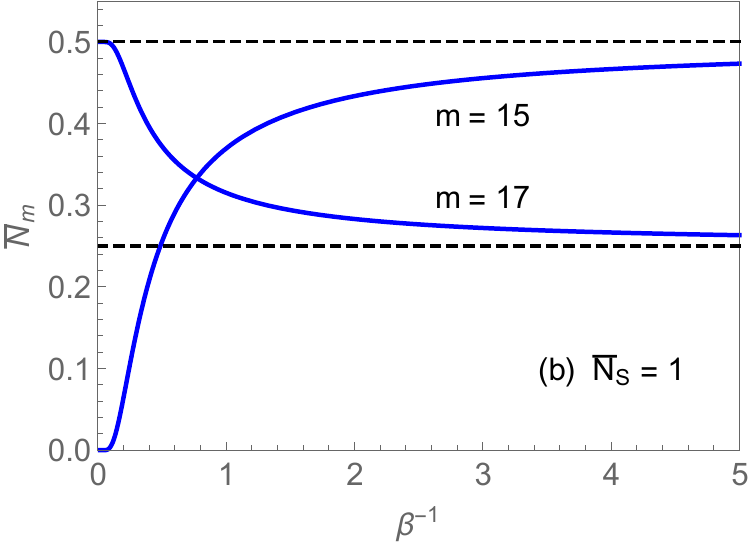}
\end{center}
\caption{Population densities (a) $\bar{N}_1$ at $\bar{N}_\mathrm{S}=0$ and (b)  $\bar{N}_{15}$, $\bar{N}_{17}$ at $\bar{N}_\mathrm{S}=1$ plotted versus $\beta^{-1}$. All other $\bar{N}_m$ vanish identically if only large or small disks are present.}
  \label{fig:A3}
\end{figure}

The cases with disks of one size can be realized with much simpler jamming protocols.
We begin with $\bar{N}_\mathrm{S}=0$ (no small disks).
The pseudo-vacuum (\ref{eq:2}) is the most compact configuration of large disks.
Particles 1 are the only species which do not include small disks. 
The population density $\bar{N}_1$ decreases monotonically and vanishes as $\beta\to\infty$ [Fig.~\ref{fig:A3}(a)].

In the case $\bar{N}_\mathrm{S}=1$, all configurations consist of particles 15 and 17 [Fig.~\ref{fig:A3}(b)] \cite{note2}.
For $\beta=0$, they are activated in a 2:1 ratio [Fig.~\ref{fig:A1}].
At reduced intensity of agitations, the macrostate frozen out by jamming contains more particles 17 and fewer particles 15.
This trend lowers the entropy and the excess volume [Fig.~\ref{fig:A2}].
For $\beta\to\infty$, the most compact configuration of small disks is realized, which has zero entropy and the minimum volume.

\begin{figure}[b]
  \begin{center}
\includegraphics[width=43mm]{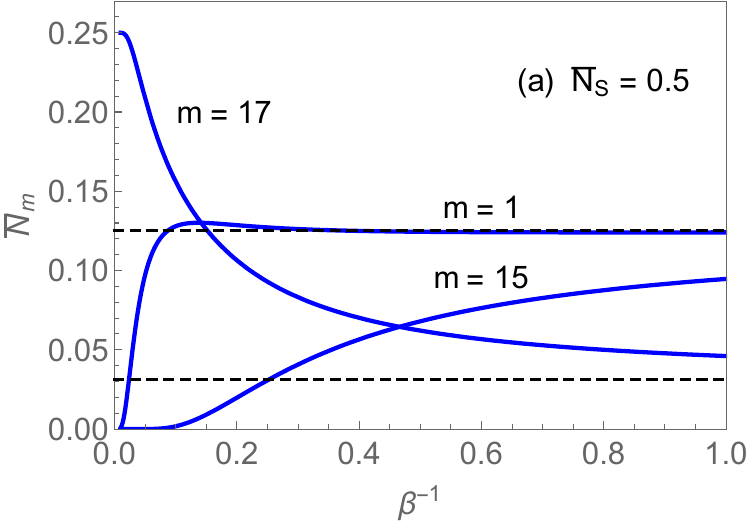}%
\hspace*{1mm}\includegraphics[width=43mm]{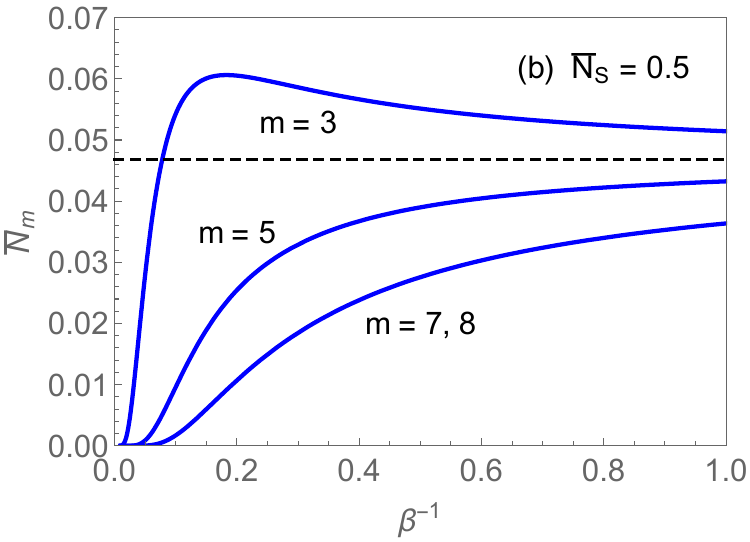}
\includegraphics[width=43mm]{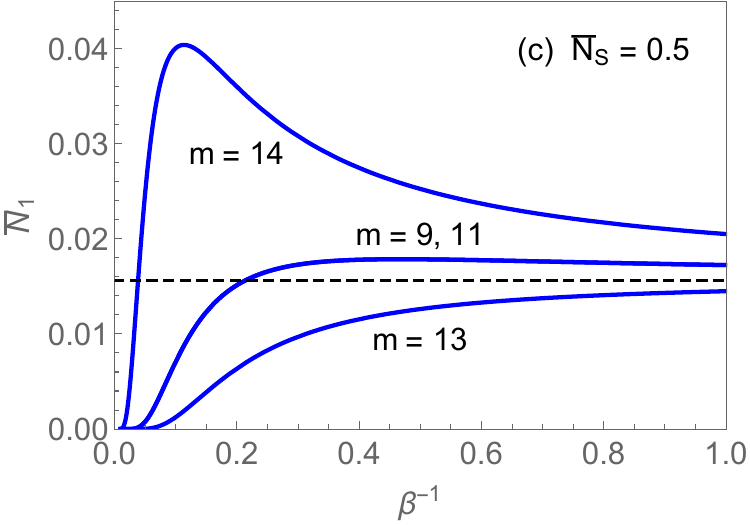}%
\hspace*{1mm}\includegraphics[width=43mm]{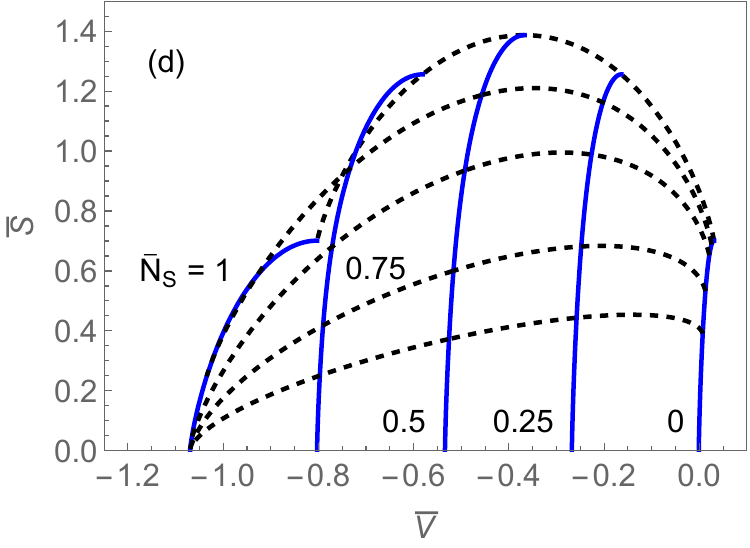}
\end{center}
\caption{Population densities $\bar{N}_m$ for particles from species  (a) $m=1,15,17$, (b)  $m=3,5,7,8$ and (c) $m=9,11,13,14$ at $\bar{N}_\mathrm{S}=0.5$ plotted versus $\beta^{-1}$.
Note the different vertical scales. In panel (d) the solid curves show $\bar{S}$ versus $\bar{V}$ parametrically for varying $\beta$ at different values of $\bar{N}_\mathrm{S}$. The dashed curves represent $\bar{S}$ versus $\bar{V}$ for vaying $\bar{N}_\mathrm{S}$ at $\beta=0,5,10,20,30$ (top to bottom). }
  \label{fig:A4}
\end{figure}

Particles from all species are activated for cases with $0<\bar{N}_\mathrm{S}<1$.
The trend toward ordering is now more complex.
We begin with the evolution of the particle population densities for the case $\bar{N}_\mathrm{S}=\frac{1}{2}$ [Figs.~\ref{fig:A4}(a)-(c)] for decreasing $\beta^{-1}$, i.e. for pre-jamming states under softening intensities of agitations.
We observe that particles 5, 7, 8, 13, 15 gradually decrease in numbers and vanish in the limit $\beta^{-1}\to0$, whereas particles 1, 3, 9, 11, 14 initially increase in numbers, go through a maximum and then also vanish.

That leaves particles 17 with a population density that keeps growing toward $\bar{N}_{17}=\frac{1}{4}$ in the limit $\beta^{-1}\to0$.
The macrostate thus evolving grows segments of reference state (representing large disks in the most compact configuration) and segments of particles 17 (representing small disks in the most compact configuration).
The segments increase in size and merge until only one of each type is left.
This is how disk segregation according to size happens in this system.

\begin{figure}[b]
  \begin{center}
\includegraphics[width=43mm]{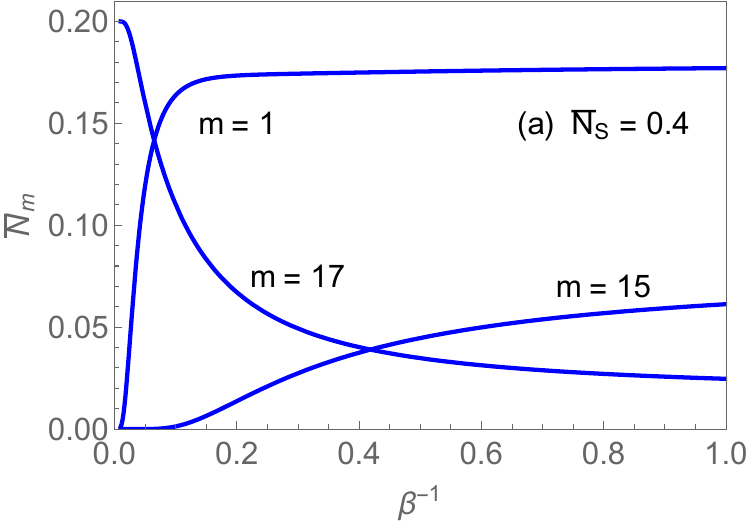}%
\hspace*{1mm}\includegraphics[width=43mm]{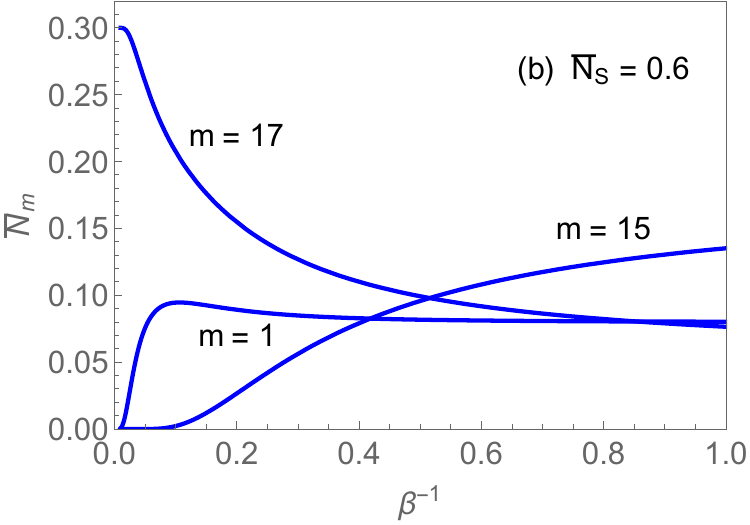}
\includegraphics[width=43mm]{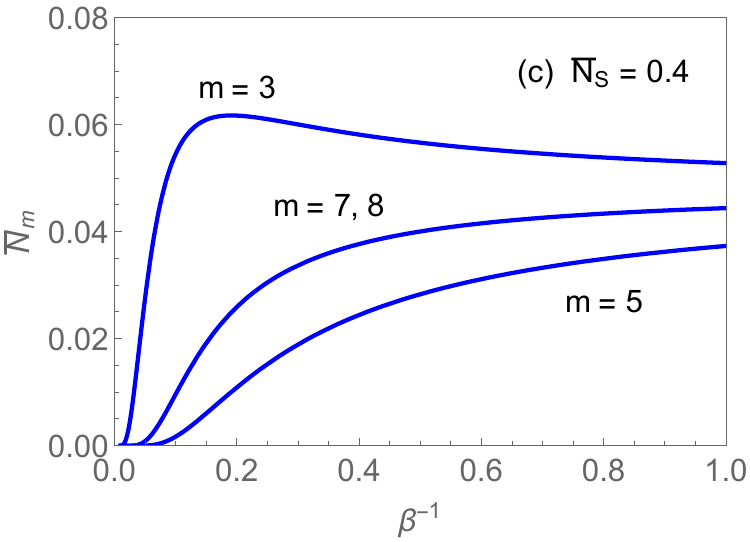}%
\hspace*{1mm}\includegraphics[width=43mm]{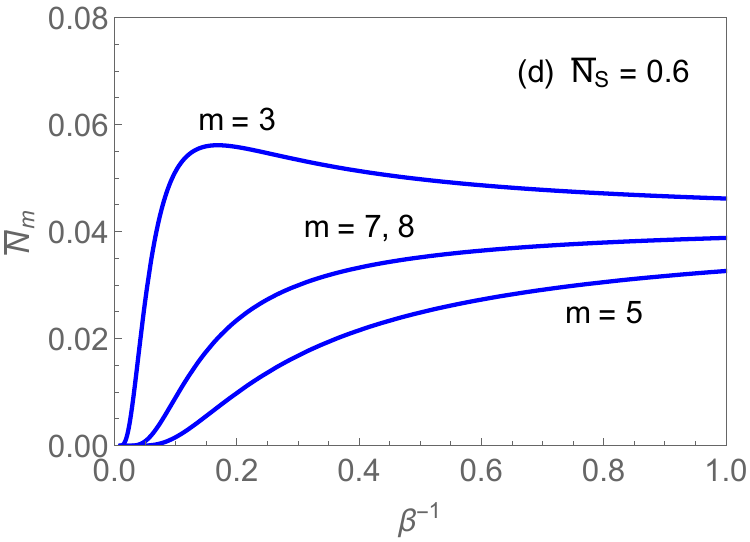}
\includegraphics[width=43mm]{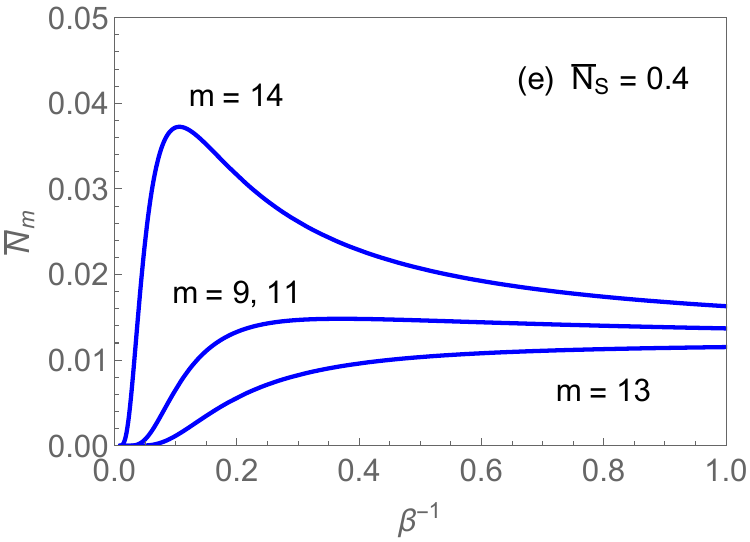}%
\hspace*{1mm}\includegraphics[width=43mm]{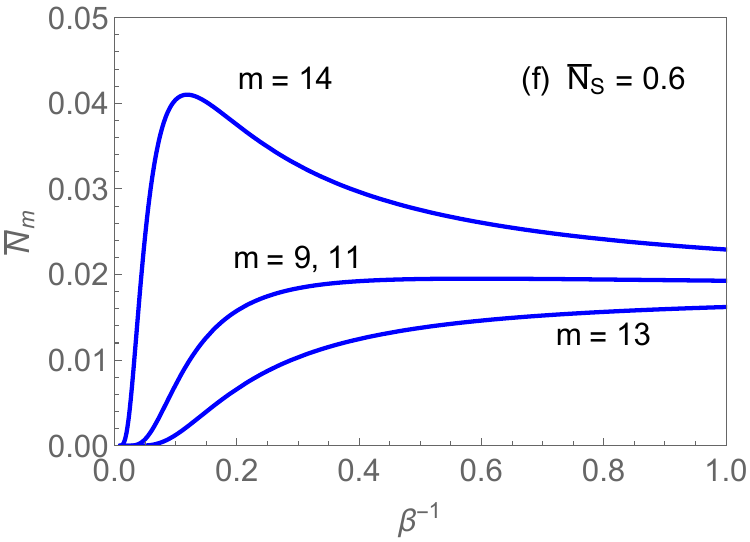}
\end{center}
\caption{Population densities $\bar{N}_m$ for particles from species all species for $\bar{N}_\mathrm{S}=0.4$ (left) and $\bar{N}_\mathrm{S}=0.6$ (right). Note the different vertical scales.}
  \label{fig:A9}
\end{figure}

The curves in Fig.~\ref{fig:A4}(d) consolidate the data used in Fig.~\ref{fig:A2} for entropy and excess volume, illuminating the ordering tendency from a different angle.
The entropy reaches zero in the limit $\beta^{-1}\to0$ for all values of $\bar{N}_\mathrm{S}$, but the approach is far from uniform. 
The excess volume for given $\bar{N}_\mathrm{S}$ changes little across the full range of $\beta$.
The truly remarkable result of this scenario is the segregation tendency between large and small disks driven by steric forces alone and without any directional bias.
 
The curves in Fig.~\ref{fig:A9}, which show results for particle population densities at $\bar{N}_\mathrm{S}=0.4$ and $\bar{N}_\mathrm{S}=0.6$ demonstrate that the trend toward size segregation persists if the fractions of small and large disks are unequal.
These graphs have benchmark value. 
We shall find that the ordering tendency realized for $\Delta\mathcal{V}<0$ assigns more complex roles to particles upon variations of $\bar{N}_\mathrm{S}$.

\subsection{Size alternation}\label{sec:ssm-b}
\begin{figure}[htb]
  \begin{center}
\includegraphics[width=43mm]{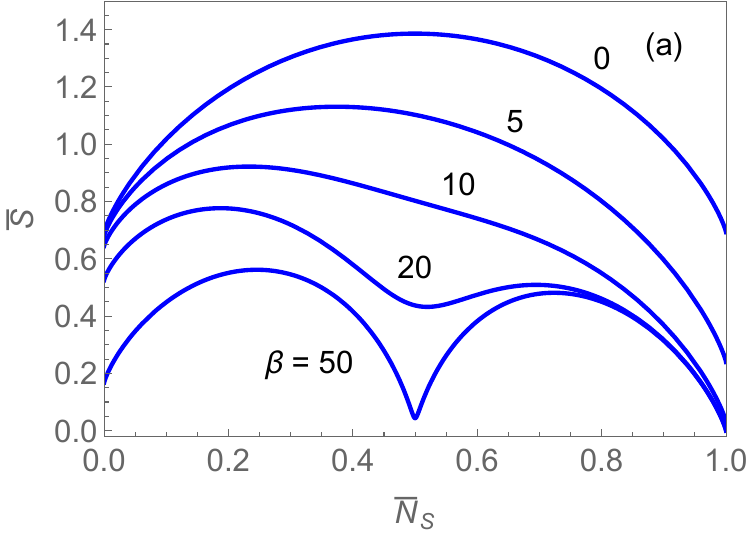}%
\hspace*{1mm}\includegraphics[width=43mm]{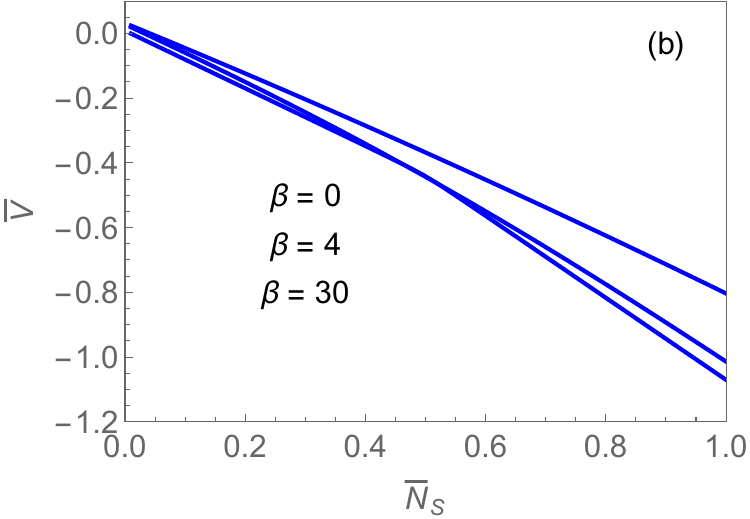}
\end{center}
\caption{(a) Entropy $\bar{S}$ and (b) excess volume $\bar{V}$ versus the fraction $\bar{N}_\mathrm{S}$ of small disks for various values of $\beta$.}
  \label{fig:A5}
\end{figure}

A representative case with $\Delta\mathcal{V}<0$ can be realized by a minor change in the energy parameter as described in Appendix~\ref{sec:appa-2}.
Nothing changes in the limit $\beta=0$ for any value of $\bar{N}_\mathrm{S}$ and in the limits $\bar{N}_\mathrm{S}=0,1$ for any value of $\beta$.
This includes expressions (\ref{eq:27}) and Figs.~\ref{fig:A1}, \ref{fig:A3}.
However, with $\beta$ increasing from zero at $0<\bar{N}_\mathrm{S}<1$, the ordering tendency now favors size alternation, which manifests itself conspicuously in the entropy, but only marginally in the excess volume when plotted versus $\bar{N}_\mathrm{S}$ [Fig.~\ref{fig:A5}].

\begin{figure}[b]
  \begin{center}
\includegraphics[width=43mm]{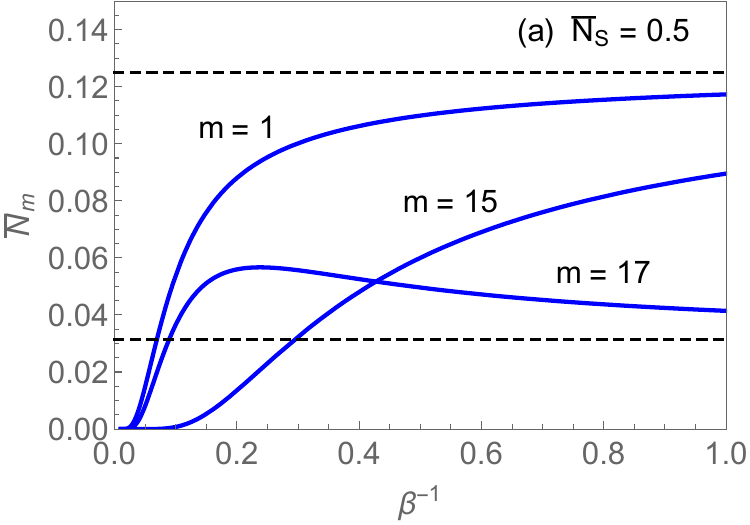}%
\hspace*{1mm}\includegraphics[width=43mm]{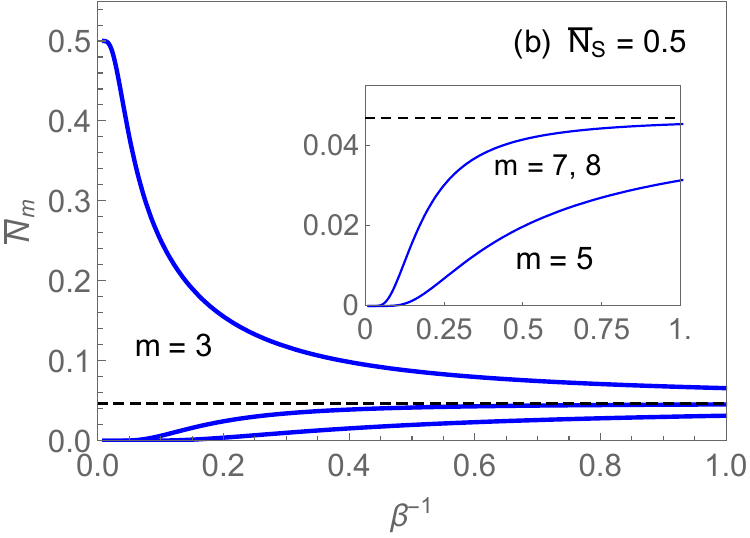}
\includegraphics[width=43mm]{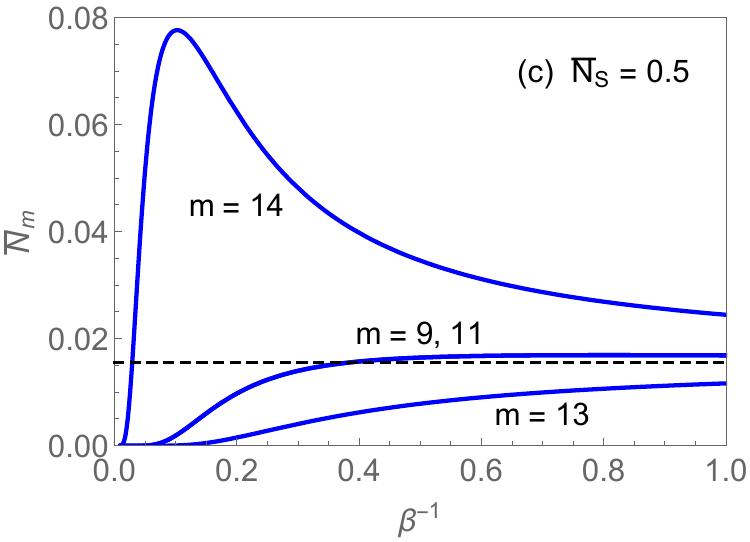}%
\hspace*{1mm}\includegraphics[width=43mm]{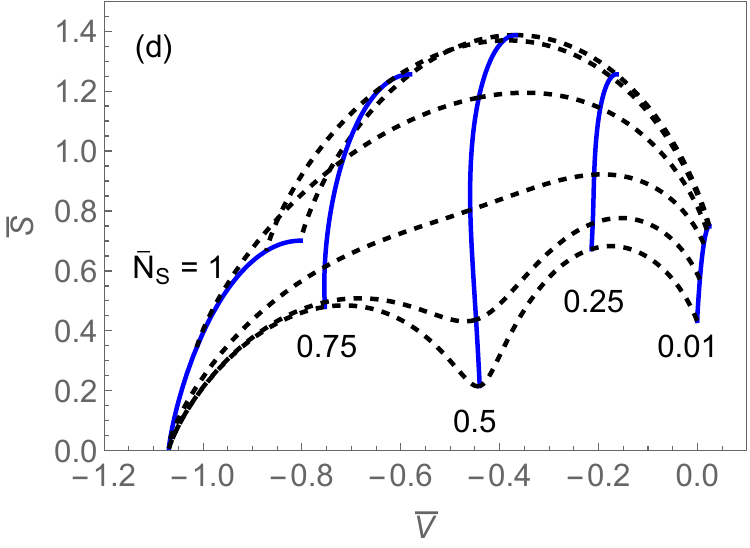}
\end{center}
\caption{Population densities $\bar{N}_m$ for particles from species  (a) $m=1,15,17$, (b)  $m=3,5,7,8$ and (c) $m=9,11,13,14$ at $\bar{N}_\mathrm{S}=1$ plotted versus $\beta^{-1}$.
Note the different vertical scales. In panel (d) the solid curves show $\bar{S}$ versus $\bar{V}$ parametrically for varying $\beta$ at different values of $\bar{N}_\mathrm{S}$. The dashed curves represent $\bar{S}$ versus $\bar{V}$ for vaying $\bar{N}_\mathrm{S}$ at $\beta=0,1,4,10,20,30$ (top to bottom). }
  \label{fig:A6}
\end{figure}

The most striking feature is that if particles of both sizes are present, the entropy dips down to zero only for the case $\bar{N}_\mathrm{S}=\frac{1}{2}$ as $\beta^{-1}\to0$.
Ordering remains incomplete for $\bar{N}_\mathrm{S}\neq\frac{1}{2}$.
Size alternation leaves a different signature than size segregation in the particle population densities.
The evidence for the case $\bar{N}_\mathrm{S}=\frac{1}{2}$ is best seen in Figs.~\ref{fig:A6}(a)-(c) when compared with Figs.~\ref{fig:A4}(a)-(c).

Here it is the particles 3 which survive in macroscopic numbers as $\beta^{-1}\to0$. 
They are the building blocks of a size-alternating sequence in the most compact configuration [Table~\ref{tab:t3}].
All other particles are frozen out gradually as $\beta^{-1}\to0$.
Some population densities pass through a smooth maximum on their way to zero as already observed on the road to size segregation [Sec.~\ref{sec:ssm-a}].

\begin{figure}[t]
  \begin{center}
\includegraphics[width=43mm]{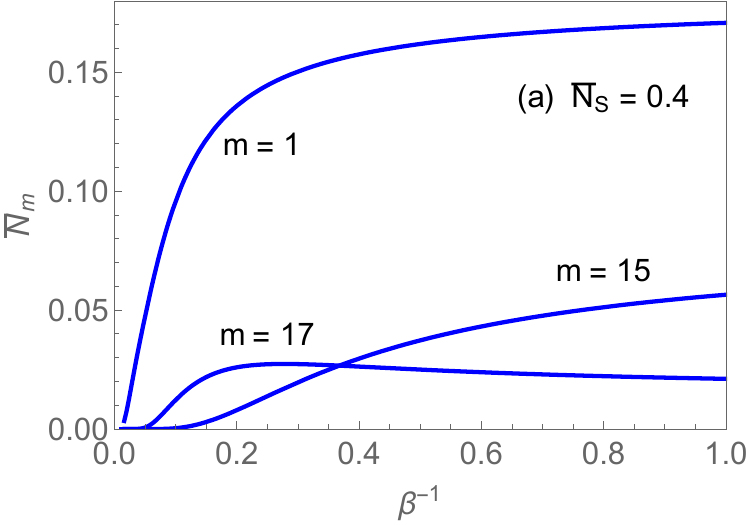}%
\hspace*{1mm}\includegraphics[width=43mm]{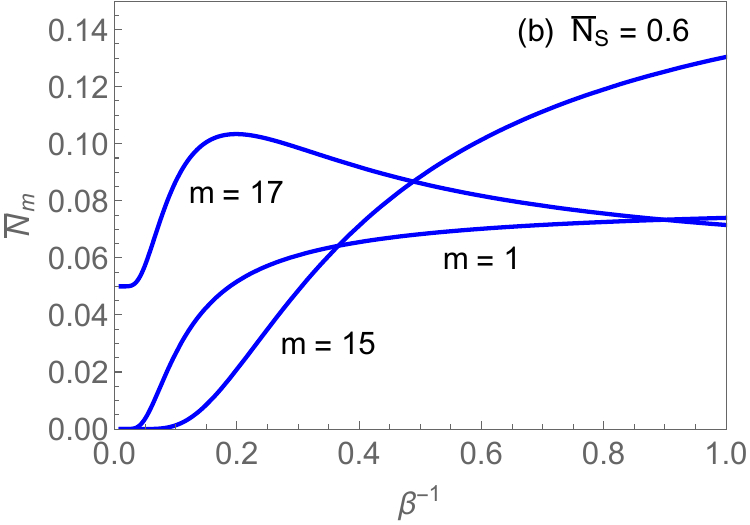}
\includegraphics[width=43mm]{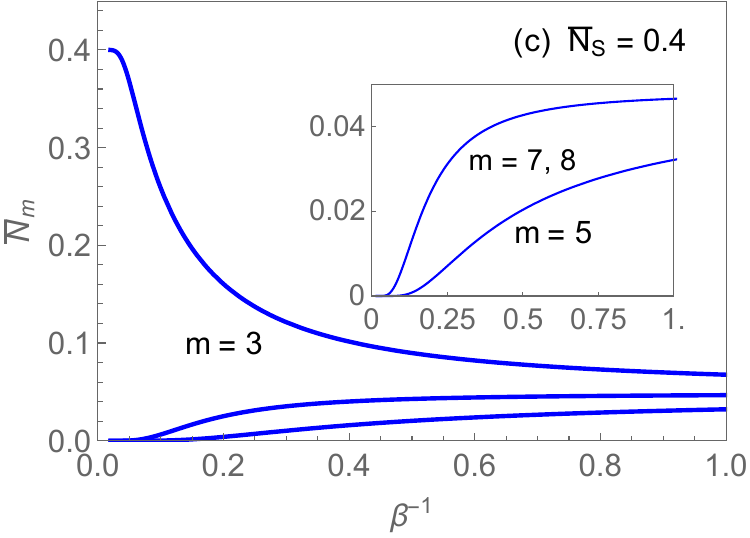}%
\hspace*{1mm}\includegraphics[width=43mm]{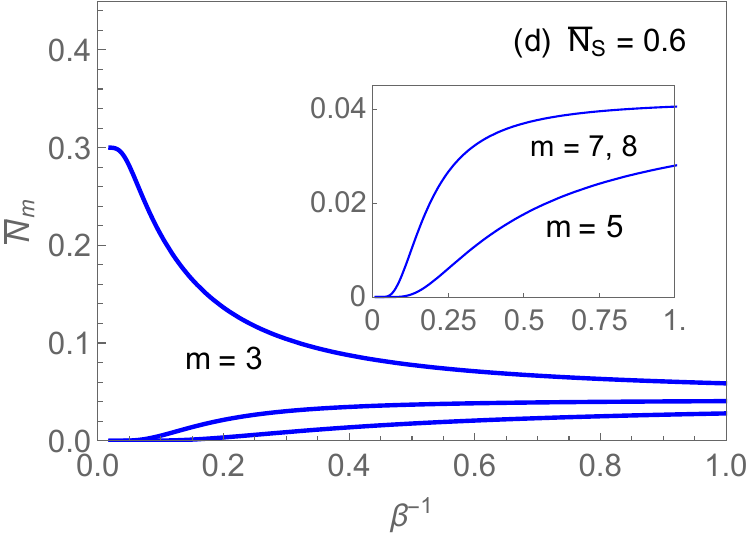}
\includegraphics[width=43mm]{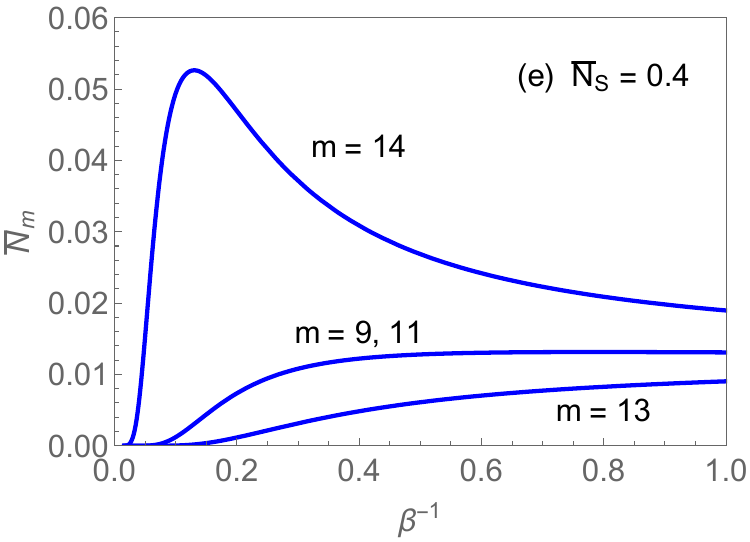}%
\hspace*{1mm}\includegraphics[width=43mm]{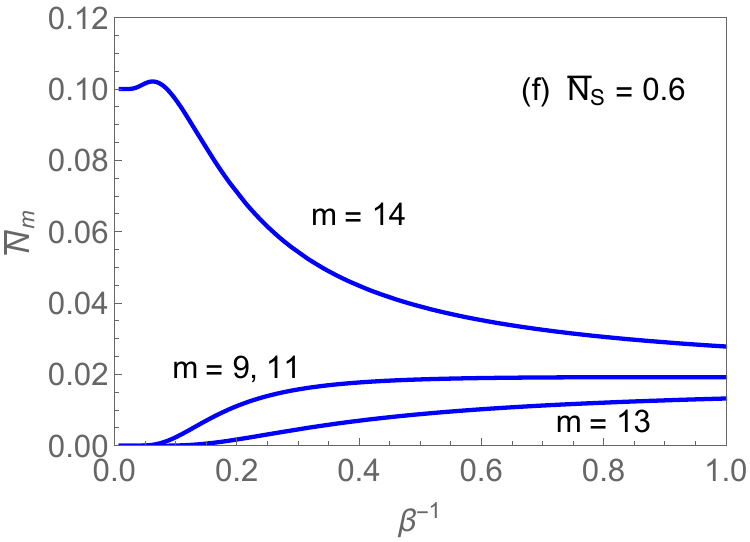}
\end{center}
\caption{Population densities $\bar{N}_m$ for particles from all species for $\bar{N}_\mathrm{S}=0.4$ (left) and $\bar{N}_\mathrm{S}=0.6$ (right). Note the different vertical scales.}
  \label{fig:A10}
\end{figure}

The common feature of the cases with $\bar{N}_\mathrm{S}=\frac{1}{2}$ leading to size segregation or size alternation is that the ordering is encoded in a single species of particles.
The two ordering tendencies play out differently for $\bar{N}_\mathrm{S}\neq\frac{1}{2}$.
Whereas size segregation can still grow to completion, size alternation must remain incomplete. 
The difference is illustrated in Fig.~\ref{fig:A10} in comparison with Fig.~\ref{fig:A9}.

At $\bar{N}_\mathrm{S}<\frac{1}{2}$, the incomplete size alternation is encoded in the diminished (still macroscopic) population of particles 3 at small $\beta^{-1}$ [Fig.~\ref{fig:A10}(c)]. 
These particles are replaced by elements of pseudo-vacuum, which consist of compactly sequenced large disks. 
The most compact jammed configuration as realized in the limit $\beta^{-1}\to0$ consists of particles 3 randomly distributed in a sea of large disks. Each particle 3 has exactly one small disk between two large disks, 
This arrangement explains the residual entropy evident in Fig.~\ref{fig:A6}(d).

At $\bar{N}_\mathrm{S}>\frac{1}{2}$ we also observe a diminished macroscopic population density $\bar{N}_3$.
However, in this instance particles 3 [Fig.~\ref{fig:A10}(d)] are replaced by particles 14 [Fig.~\ref{fig:A10}(f)] and 17 [Fig.~\ref{fig:A10}(b)]. 
The excess numbers of small disks are being hosted by host 3 and host 14 in the form of tags 17. 
Each tag 17 adds two small disks to its host.
Hence a hosting particle 3 contains an odd number of small disks and a hosting particle 14 an even number of small disks.
Unlike at $\bar{N}_\mathrm{S}<\frac{1}{2}$, where hosts 3 were randomly distributed in the pseudo-vacuum, here the hosts 3 and 14 are close packed.
The residual entropy, evident in Fig.~\ref{fig:A6}(d), is associated with two types of hosts and their tag contents.

\begin{figure}[t]
  \begin{center}
\includegraphics[width=60mm]{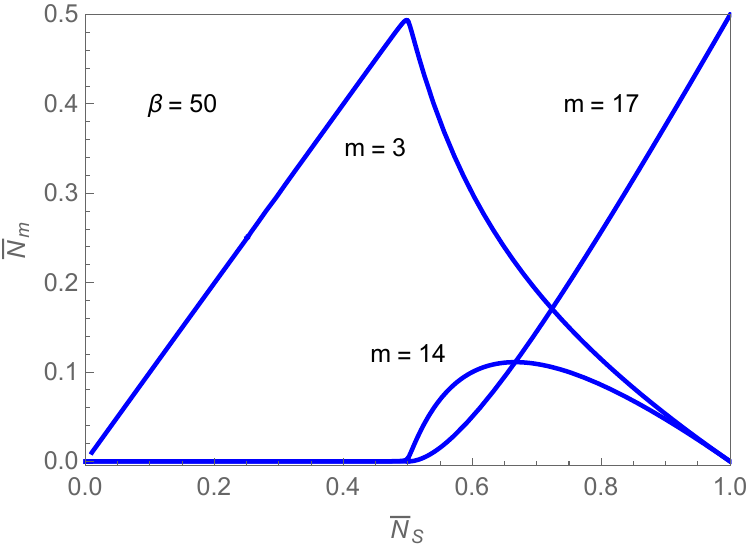}
\end{center}
\caption{Population densities $\bar{N}_m$ versus $\bar{N}_\mathrm{S}$  at $\beta=50$ for particles present in macroscopic numbers in the limit $\beta\to\infty$.}
  \label{fig:A12}
\end{figure}

The shifting composition of the most compact disk configuration with $\bar{N}_\mathrm{S}$ varying across its full range is illustrated in Fig.~\ref{fig:A12}.
At $\bar{N}_\mathrm{S}=0$ the most compact state is the pseudo-vacuum (devoid of any particles).
Every small disk added is embedded in a host 3 and these particles are randomly distributed in the pseudo-vacuum.
The entropy first rises and then becomes lower again as the space gets crowded for particles 3 [{Fig.~\ref{fig:A5}]. 
At $\bar{N}_\mathrm{S}=\frac{1}{2}$ particles 3 are packed solid with the disks fully ordered in a size alternating pattern. 
As the small disks become the majority, the population of tags 17 takes off.
They are being hosted by particles 3 in random odd numbers. 
The population density of particles 14, which are alternate hosts for even numbers of tags 17, also grows from zero.
The diminishing number of large disks ultimately suppresses the populations of both hosts back toward zero. 
At $\bar{N}_\mathrm{S}=1$, only one host 3 or 14 will be left, which contains all tags 17.

\begin{figure}[t]
  \begin{center}
\includegraphics[width=43mm]{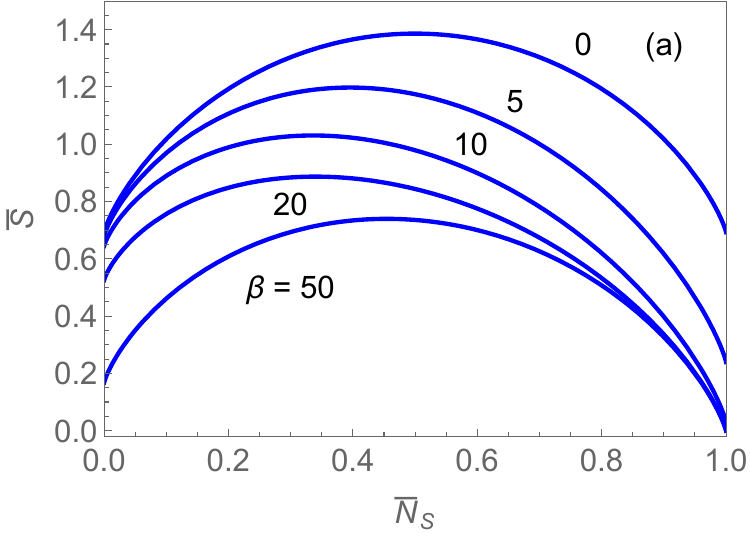}%
\hspace*{1mm}\includegraphics[width=43mm]{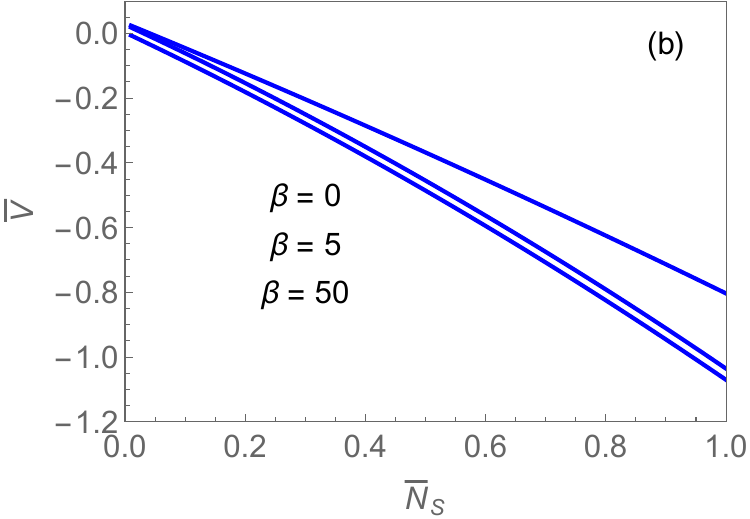}
\end{center}
\caption{(a) Entropy $\bar{S}$ and (b) excess volume $\bar{V}$ versus the fraction $\bar{N}_\mathrm{S}$ of small disks  for various values of $\beta$.}
  \label{fig:A7}
\end{figure}

\begin{figure}[b]
  \begin{center}
\includegraphics[width=43mm]{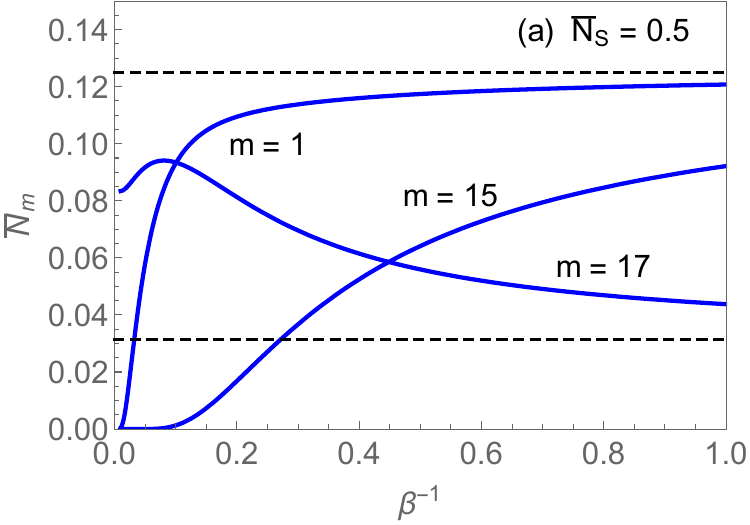}%
\hspace*{1mm}\includegraphics[width=43mm]{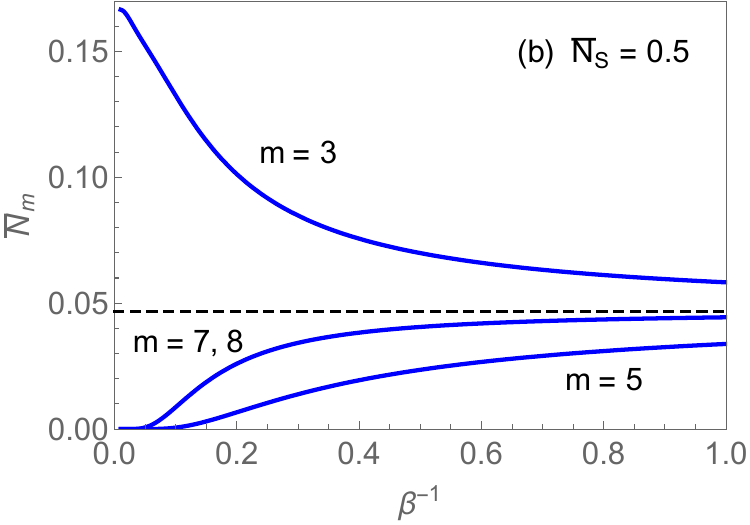}
\includegraphics[width=43mm]{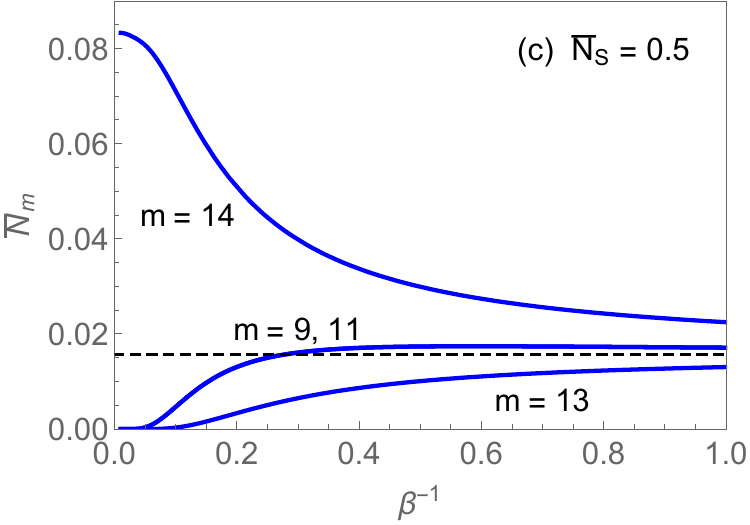}%
\hspace*{1mm}\includegraphics[width=43mm]{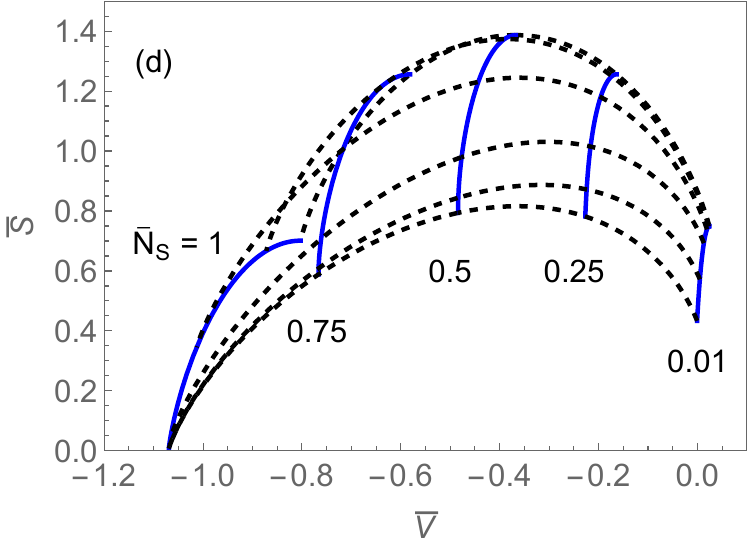}
\end{center}
\caption{Population densities $\bar{N}_m$ for particles from species  (a) $m=1,15,17$, (b)  $m=3,5,7,8$ and (c) $m=9,11,13,14$ at $\bar{N}_\mathrm{S}=1$ plotted versus $\beta^{-1}$.
Note the different verical scales. In panel (d) the solid curves show $\bar{S}$ versus $\bar{V}$ parametrically for varying $\beta$ at different values of $\bar{N}_\mathrm{S}$. The dashed curves represent $\bar{S}$ versus $\bar{V}$ for vaying $\bar{N}_\mathrm{S}$ at $\beta=0,1,4,10,20,30$ (top to bottom). }
  \label{fig:A8}
\end{figure}

\subsection{Size randomness}\label{sec:ssm-c}
It is instructive to consider a case at the border between the two regimes with ordering tendencies toward size segregation on one side and size alternation on the other.
This border case, associated with $\Delta\mathcal{V}=0$, exhibits persistent size randomness in the limit $\beta^{-1}\to0$ for $0<\bar{N}_\mathrm{S}<1$.
The exact analysis produces its own characteristic features.

We showed earlier that for $\bar{N}_\mathrm{S}=\frac{1}{2}$ both size segregation and size alternation are established to completion in the limit $\beta^{-1}\to0$, whereas for $\bar{N}_\mathrm{S}\neq\frac{1}{2}$ only size segregation still reaches completion.
What distinguishes the borderline case is that even though the macrostate shows evidence of compactification in the limit $\beta^{-1}\to0$, the disk sequence remains random for $0<\bar{N}_\mathrm{S}<1$, giving rise to a persistent entropy [Fig.~\ref{fig:A7}].

\begin{figure}[t]
  \begin{center}
\includegraphics[width=43mm]{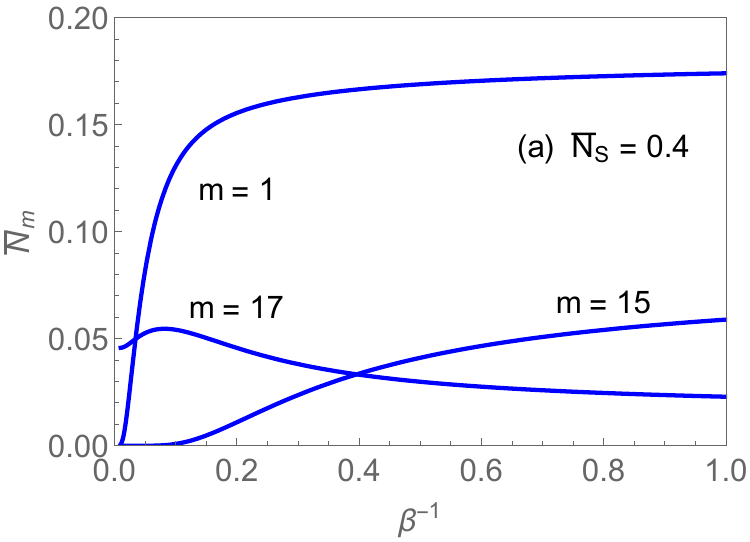}%
\hspace*{1mm}\includegraphics[width=43mm]{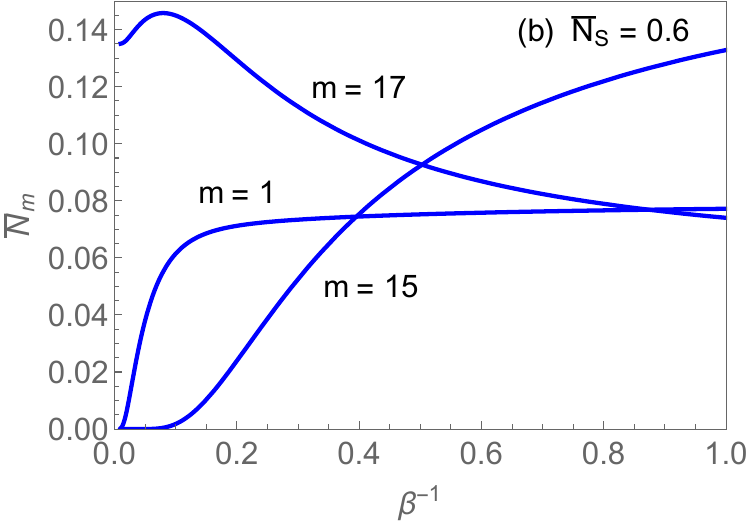}
\includegraphics[width=43mm]{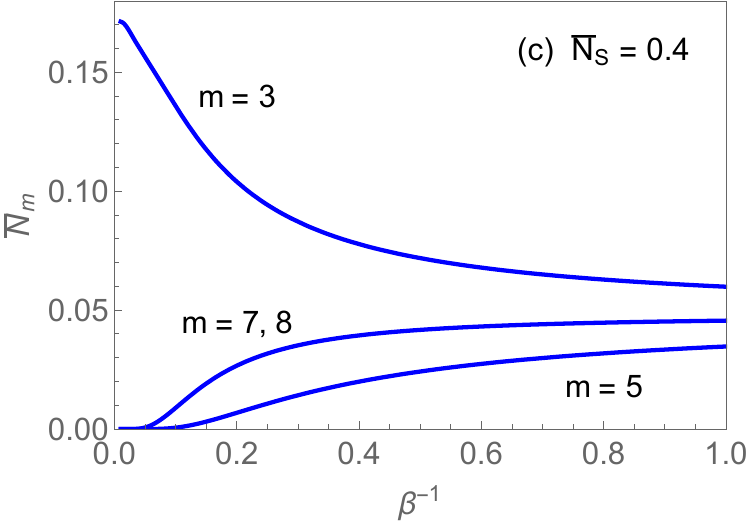}%
\hspace*{1mm}\includegraphics[width=43mm]{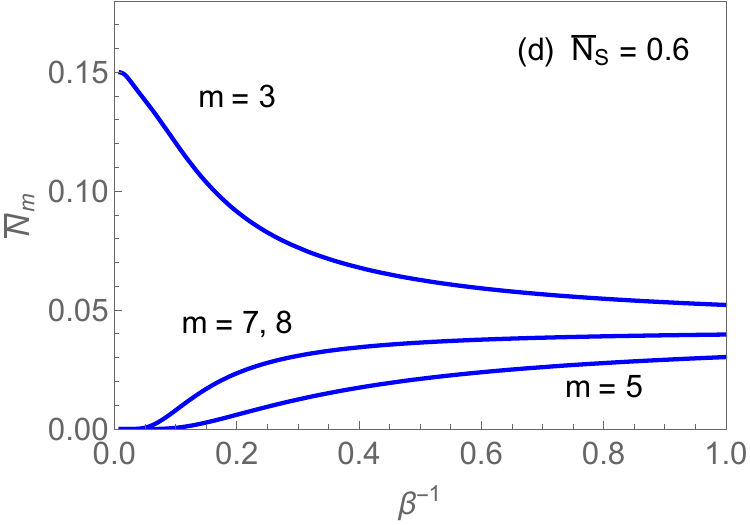}
\includegraphics[width=43mm]{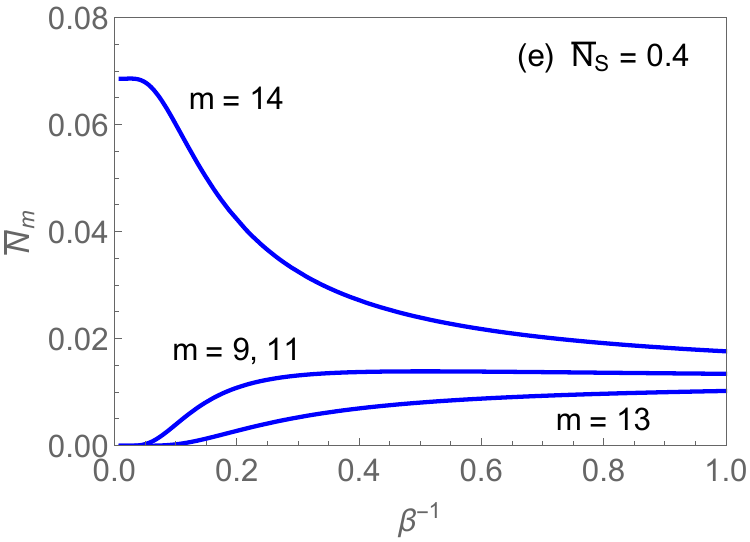}%
\hspace*{1mm}\includegraphics[width=43mm]{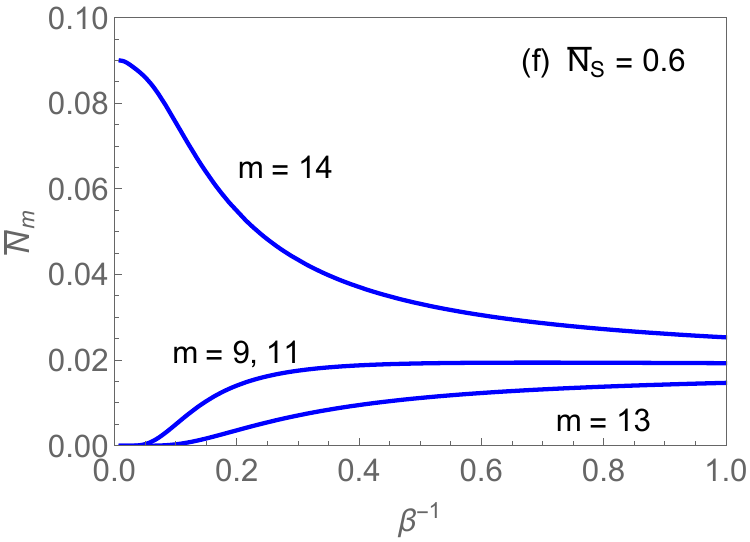}
\end{center}
\caption{Population densities $\bar{N}_m$ for particles from species all species for $\bar{N}_\mathrm{S}=0.4$ (left) and $\bar{N}_\mathrm{S}=0.6$ (right). Note the different vertical scales.}
  \label{fig:A11}
\end{figure}

In Fig.~\ref{fig:A8}(a)-(c) the focus is on $\bar{N}_\mathrm{S}=\frac{1}{2}$ for direct comparison with the corresponding data in Figs.~\ref{fig:A4} and \ref{fig:A6}.
The compactification is caused by the disappearance of particles from all species that have successive disks touching the same wall.
This leaves only particles 3, 14, and 17. 
The jammed macrostate in the limit $\beta^{-1}\to0$ consists of randomly distributed hosts 3 and 14, each containing a random number of tags 17.
The only constraint is that small and large disks are present in equal numbers.

When we vary $\bar{N}_\mathrm{S}$ away from $\frac{1}{2}$, no qualitative difference appears [Fig.~\ref{fig:A11}]. 
The robustness of disorder is akin to the robustness of order in the form of size segregation [Fig.~\ref{fig:A9}]. 
The order in the form of size alternation lacks this trait [Fig.~\ref{fig:A10}].
It remains incomplete except for $\bar{N}_\mathrm{S}=0,\frac{1}{2},1$.
The contrasting ordering patterns in the three cases are best compared in panel (d) of Figs.~\ref{fig:A4},\ref{fig:A6},\ref{fig:A8}.

\begin{figure}[b]
  \begin{center}
\includegraphics[width=60mm]{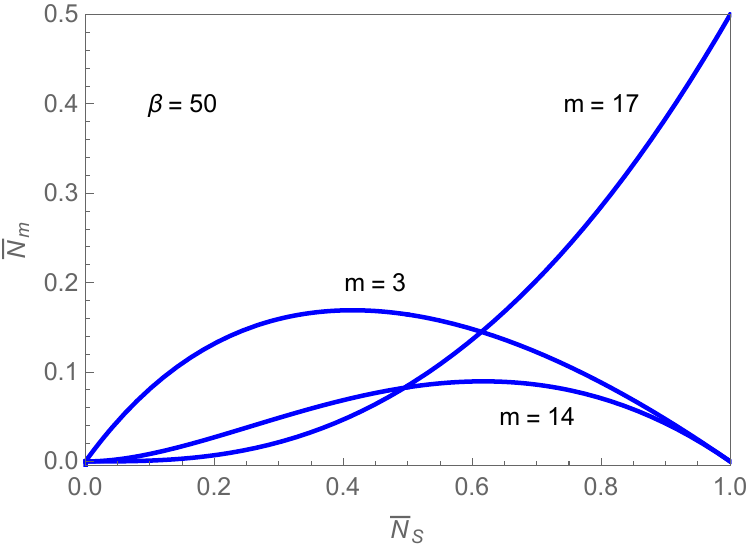}
\end{center}
\caption{Population densities $\bar{N}_m$ versus $\bar{N}_\mathrm{S}$  at $\beta=50$ for particles present in macroscopic numbers in the limit $\beta\to\infty$.}
  \label{fig:A13}
\end{figure}

For the border case, the population densities of the surviving particles 3, 14, and 17 for large $\beta$ as plotted versus $\bar{N}_\mathrm{S}$ in Fig.~\ref{fig:A13} show the competing ordering and disordering tendencies in play. 
The ordering tendency at large $\beta$ has frozen out significant populations of all other particle species. 
The lack of preference among the survivors as enforced by the balanced energy parameters $(\Delta\mathcal{V}=0)$ establishes population densities which maximize the entropy under these constraints. 

Hosts 3 dominate hosts 14 at small $\bar{N}_\mathrm{S}$ for the simple reason that (without hosting any tags 17) they embed only one small disk as opposed to two. 
The scarcity of small disks can be most effectively spread in a random fashion by hosts 3.
For large $\bar{N}_\mathrm{S}$, on the other hand, when both hosts are loaded with tags, it makes little difference if the number of small disks  carried by the host is even (host 14) or odd (host 3). 
Both hosts have almost indistinguishable population densities, much smaller than the population density of tags 17.

%
\section{Emphasis on stability}\label{sec:scen2}
%
We stated earlier that the model derived from the $s=\frac{3}{2}$ Ising model included some jammed disk configurations that are not stable.
This vulnerability was the price to be paid to maintain the symmetries that made the mathematical analysis simple and transparent.
Here and in Appendix~\ref{sec:appb} we consider a model which only permits jammed disk configurations that are mechanically stable.
Its analysis requires an extended computational component, which renders it less transparent, but it produces results with the exact same ordering tendencies.

We noted earlier that particles $m=1,2,15,16$ in Tables \ref{tab:t3} and \ref{tab:t7} are only marginally stable, whereas particles $m=5,6$ are unstable in some (all) configurations when activated from reference state $\mathsf{pv_L}$ ($\mathsf{pv_S}$).
Full mechanical stability for all jammed microstates can be restored by modifications in the statistical interactions between particles as highlighted in Table \ref{tab:t8}:

\begin{itemize}

\item[--] Eliminating rows/columns $5,6$ from Table \ref{tab:t4} prevents the activation of hosts $m=5,6$.

\item[--] The marginal stability of particles $m=1,2,15,16$ is upgraded to full stability if we restrict their placement options.

\item[--] Compacts $m=1,2$ are fully stable unless two or more of the same are positioned back to back. Such occurrences are eliminated if we modify $g_{11}$ and $g_{22}$ as shown.

\item[--] Tag $m=15$ ($m=16$) must be prevented from being hosted by hosts $m=7,9,11,13$ ($m=8,10,12,14$) to ensure mechanical stability. 
This requires modifications of $g_{15,7},g_{15,9},g_{15,11},g_{15,13}$ ($g_{16,8},g_{16,10},g_{16,12},g_{16,14}$).

\item[--] When tag $m=15$ ($m=16$) is hosted by hosts $m=4,10,12,14$ ($m=3,9,11,13$), it must be prevented from multiple occupancy in the same host, which requires the modification of $g_{15,16}$  and $g_{16,15}$. 

\end{itemize}

\begin{table}[t]
  \caption{Statistical interaction coefficients of 15 species of particles with motifs compiled in Table~\ref{tab:t7}. These particles generate all stable jammed microstates from the reference state (\ref{eq:36}). The capacity constants $A_m$ and quantum numbers $s_m$ are as in Table~\ref{tab:t4}. The modified $g_{mn}$ are highlighted.}\label{tab:t8} 
\begin{center}
\begin{tabular}{r|rrrrrrrrrrrrrrr} 
$g_{mn}$ & $~1$ & $~2$ & $3$ & $4$ & $7$ & $8$ & $9$ & $10$ & 
$11$ & $12$ & $13$ & $14$ & $~15$ & $~16$ & $17$ \\
\hline \rule[-2mm]{0mm}{6mm}
$1$ & $\mathbf{\frac{3}{2}}$ & $\frac{1}{2}$ & $0$ & $1$ 
& $\frac{1}{2}$ & $\frac{1}{2}$ 
& $\frac{1}{2}$ & $\frac{3}{2}$ & $\frac{1}{2}$ & $\frac{3}{2}$ 
& $1$ & $1$ & $\frac{1}{2}$ & $\frac{1}{2}$ 
& $1$ \\ \rule[-2mm]{0mm}{5mm}
$2$ & $\frac{1}{2}$ & $\mathbf{\frac{3}{2}}$ & $1$ & $0$ 
& $\frac{1}{2}$ & $\frac{1}{2}$ 
& $\frac{3}{2}$ & $\frac{1}{2}$ & $\frac{3}{2}$ & $\frac{1}{2}$ 
& $1$ & $1$ & $\frac{1}{2}$ & $\frac{1}{2}$ 
& $1$ \\ \rule[-2mm]{0mm}{5mm}
$3$ & $\frac{1}{2}$ & $\frac{1}{2}$ & $1$ & $1$ 
& $\frac{1}{2}$ & $\frac{1}{2}$ 
& $\frac{1}{2}$ & $\frac{3}{2}$ & $\frac{1}{2}$ & $\frac{3}{2}$ 
& $1$ & $1$ & $\frac{1}{2}$ & $\frac{1}{2}$ 
& $1$ \\ \rule[-2mm]{0mm}{5mm}
$4$ & $\frac{1}{2}$ & $\frac{1}{2}$ & $1$ & $1$ 
& $\frac{1}{2}$ & $\frac{1}{2}$ 
& $\frac{3}{2}$ & $\frac{1}{2}$ & $\frac{3}{2}$ & $\frac{1}{2}$ 
& $1$ & $1$ & $\frac{1}{2}$ & $\frac{1}{2}$ 
& $1$ \\ \rule[-2mm]{0mm}{5mm}
$7$ & $1$ & $1$ & $2$ & $2$ 
& $2$ & $1$ 
& $3$ & $3$ & $3$ & $3$ 
& $3$ & $3$ & $1$ & $1$ 
& $2$ \\ \rule[-2mm]{0mm}{5mm}
$8$ & $1$ & $1$ & $2$ & $2$ 
& $2$ & $2$ 
& $3$ & $3$ & $3$ & $3$ 
& $3$ & $3$ & $1$ & $1$ 
& $2$ \\ \rule[-2mm]{0mm}{5mm}
$9$ & $\frac{1}{2}$ & $\frac{1}{2}$ & $1$ & $1$ 
& $\frac{1}{2}$ & $\frac{1}{2}$ 
& $\frac{3}{2}$ & $\frac{3}{2}$ & $\frac{1}{2}$ & $\frac{3}{2}$ 
& $1$ & $1$ & $\frac{1}{2}$ & $\frac{1}{2}$ 
& $1$ \\ \rule[-2mm]{0mm}{5mm}
$10$ & $\frac{1}{2}$ & $\frac{1}{2}$ & $1$ & $1$ 
& $\frac{1}{2}$ & $\frac{1}{2}$ 
& $\frac{3}{2}$ & $\frac{3}{2}$ & $\frac{3}{2}$ & $\frac{1}{2}$ 
& $1$ & $1$ & $\frac{1}{2}$ & $\frac{1}{2}$ 
& $1$ \\ \rule[-2mm]{0mm}{5mm}
$11$ & $\frac{1}{2}$ & $\frac{1}{2}$ & $1$ & $1$ 
& $\frac{1}{2}$ & $\frac{1}{2}$ 
& $\frac{3}{2}$ & $\frac{3}{2}$ & $\frac{3}{2}$ & $\frac{3}{2}$ 
& $1$ & $1$ & $\frac{1}{2}$ & $\frac{1}{2}$ 
& $1$ \\ \rule[-2mm]{0mm}{5mm}
$12$ & $\frac{1}{2}$ & $\frac{1}{2}$ & $1$ & $1$ 
& $\frac{1}{2}$ & $\frac{1}{2}$ 
& $\frac{3}{2}$ & $\frac{3}{2}$ & $\frac{3}{2}$ & $\frac{3}{2}$ 
& $1$ & $1$ & $\frac{1}{2}$ & $\frac{1}{2}$ 
& $1$ \\ \rule[-2mm]{0mm}{5mm}
$13$ & $1$ & $1$ & $2$ & $2$ 
& $1$ & $1$ 
& $3$ & $3$ & $3$ & $3$ 
& $3$ & $2$ & $1$ & $1$ 
& $2$ \\ \rule[-2mm]{0mm}{5mm}
$14$ & $1$ & $1$ & $2$ & $2$ 
& $1$ & $1$ 
& $3$ & $3$ & $3$ & $3$ 
& $3$ & $3$ & $1$ & $1$ 
& $2$ \\ \rule[-2mm]{0mm}{5mm}
$15$ & $0$ & $0$ & $0$ & $-1$ 
& $\mathbf{0}$ & $0$ 
& $\mathbf{0}$ & $-1$ & $\mathbf{0}$ & $-1$ 
& $\mathbf{0}$ & $-1$ & $\mathbf{1}$ & $0$ 
& $-1$ \\ \rule[-2mm]{0mm}{5mm}
$16$ & $0$ & $0$ & $-1$ & $0$ 
& $0$ & $\mathbf{0}$ 
& $-1$ & $\mathbf{0}$ & $-1$ & $\mathbf{0}$ 
& $-1$ & $\mathbf{0}$ & $0$ & $\mathbf{1}$ 
& $-1$ \\ \rule[-2mm]{0mm}{5mm}
$17$ & $0$ & $0$ & $-1$ & $-1$ 
& $-1$ & $-1$ 
& $-1$ & $-1$ & $-1$ & $-1$ 
& $-1$ & $-1$ & $0$ & $0$ 
& $0$ \\ 
\end{tabular}
\end{center}
\end{table} 

The last intervention switches particles $m=15,16$ from the category of tags to the category of \emph{caps}. 
Whereas a host can accommodate any number of tags, it can host only one cap.
The capacity constants $A_m$ and energy parameters $\bar{e}_m$ are unaffected.

The mathematical analysis with these specifications is reported in Appendix~\ref{sec:appb}.
The lower symmetry of this model puts an explicit analytic solution out of reach.
However, it is possible to always identify the (unique) physically relevant solution among the roots of a sixth-order polynomial equation.

The key evidence for the equivalence of the more stable model and the more symmetric model regarding ordering tendencies comes from an expansion of the partition function $\bar{Z}$ for large $\beta$, i.e. for soft random agitations prior to jamming.
Not only does the border between the regimes of size segregation and size alternation remain exactly the same, the large-$\beta$ expansion of $1/\bar{Z}$ for the two models only differ in the $2^\mathrm{nd}$-order correction as demonstrated in Eqs.~(\ref{eq:B16})-(\ref{eq:B18}).
Separate expansions pertain to the two ordering regimes in both models. The limits toward the border case, where size randomness persists, are equivalent from either side for both models.

In summary, there is no need to present plots for results of the model which eliminates jammed microstates of marginal mechanical stability.
The curves do show deviations for small and moderate $\beta$. 
None are of a qualitative nature and they gradually disappear at large $\beta$, where the ordering tendencies are in strong evidence.
Particles 5,6 in Table~\ref{tab:t7}, which are unstable in any jammed configuration, are removed altogether. 
For particles 1,2,15,16 in Tables~\ref{tab:t3} and \ref{tab:t7}, which are only unstable or marginally stable in some configurations, we modified the statistical interactions such that only stable configurations are retained (see Table~\ref{tab:t8}).

%
\section{CONCLUSION}\label{sec:concl}
%
This work sheds light on an important question in granular physics:
Is it possible to observe the phenomenon of segregation according to grain size in situations with no bias provided by (i) external fields, (ii) particular shapes of container walls, or (iii) discriminating interactions between grains? 

No claim is made for the reported diverse ordering tendencies that it is a spontaneously broken symmetry that leads to them. 
Recall that our rigorous analysis leaves a set of energy parameters undetermined.
The values of these energy parameters depend on the jamming protocol in use.
It is the vagaries associated with jamming protocols that limit the predictions of order in granular statistics.
In strong contrast, equilibrium statistical mechanics is free of any such caveat.

The scenario for which we have demonstrated the possibility of two distinct ordering tendencies in the absence of the biases (i)-(iii) is admittedly very special, but it will prove difficult to extend it to more complex situations that permit a rigorous analysis in the framework of configurational statistics.

The grains have been modeled as disks of two sizes confined to a narrow channel and allowed to mix before being jammed. 
The only interactions are provided by steric forces free of any bias.
The unjammed state under random agitations of given intensity is specified by a set of energy parameters whose values depend on the details of the jamming protocol in operation.

The analysis shows that the ordering tendency in the configuration of jammed disks hinges on just one inequality among the energy parameters.
The parameter space which is most likely accessible to jamming protocols divides into two regions, one favoring size segregation, the other size alternation, and the border between the regions size randomness.

The exact analysis is based on a method invented in quantum statistics \cite{Hald91a, Wu94} and adapted to the configurational statistics of quasiparticles constructed from all possible two-disk tiles \cite{janac1, janac2, janac3}.
This analysis produces exact expressions for the configurational entropy and the excess volume (relative to the most compact configuration) and other quantities as functions of the fraction of small (or large) disks and a measure of the intensity of random fluctuations prior to jamming.

\appendix

%
\section{MODEL OF HIGHER SYMMETRY}\label{sec:appa}
%
The exact analysis of the jammed macrostates for the model with emphasis on symmetry (Sec.~\ref{sec:scen1}) is outlined here.
The partition function (\ref{eq:18}) can be simplified into
\begin{align}\label{eq:a16}
  \bar{Z}^{-1}= \zeta = \prod_{m}x_{m}^{\bar{A}_{m}}=\prod_{m<15}x_m,
\end{align}
with the $x_m$ defined in (\ref{eq:17}) and for the $\bar{A}_m$, $\hat{g}_{mn}$ in Table~\ref{tab:t5}.
A corresponding simplification occurs in Eqs.~(\ref{eq:19}) for the variable $x_1$:
\begin{equation}\label{eq:a17}
  x_{1} = 1 -e_{1}\prod_nx_{n}^{\hat{g}_{n1}}
  = 1 -\bar{e}_{1}\prod_{n<15}x_{n}.
\end{equation}
In consequence, the partition function reduces to a function of $x_1$ alone:
\begin{equation}\label{eq:a9}
\bar{Z}=\frac{\bar{e}_1}{1-x_1}.
\end{equation}
For the following we introduce the fugacity $z$ and the modified energy parameters $\bar{e}_m$:
\begin{equation}\label{eq:a5}
z\doteq e^{\beta\mu},\quad
\bar{e}_m\doteq e^{-\beta p_m}=e_mz^{-s_m}.
\end{equation}
For the purpose of compact expressions and computational efficiency it is useful to introduce
\begin{align}\label{eq:a61}
y_m\doteq 1-x_m.
\end{align}

By judiciously sequencing the remaining Eqs.~(\ref{eq:19}) we can relate all $x_m$ to $\zeta$ directly or recursively:
\begin{equation}\label{eq:a3}
x_{1} = 1 -\bar{e}_{1}\zeta,
\end{equation}
\begin{equation}\label{eq:a18}
 \frac{y_{15}}{y_{1}} =\frac{e_{15}}{e_{1}}
 \quad \Rightarrow~ 
 x_{15} = 1-\bar{e}_{15}z\zeta,
\end{equation}
\begin{equation} \label{eq:a19}
  \frac{y_{17}}{y_{15}^{2}}=\frac{e_{17}}{e_{15}^{2}}x_{15}^{-2}
   \quad \Rightarrow~
  x_{17} = 1 - \frac{\bar{e}_{17}z^2\zeta^{2}}{(1-\bar{e}_{15}z\zeta)^{2}},
\end{equation}
\begin{equation}\label{eq:a20}
  \frac{y_{8}}{y_{15}}=\frac{e_{8}}{e_{15}}\frac{x_{8}}{x_{15}x_{17}}
   \quad \Rightarrow~ 
  x_{8} =  \left[1+\frac{\bar{e}_{8}z\zeta}{x_{15}x_{17}}\right]^{-1},
\end{equation}
\begin{equation}\label{eq:a21}
  \frac{y_{7}}{y_{8}}=x_{7}
   \quad \Rightarrow~   
  x_{7} = \frac{1}{2-x_{8}},
\end{equation}
\begin{equation}\label{eq:a22}
  \frac{y_{5}}{y_{17}}=\frac{e_{5}}{e_{17}}\frac{x_{15}}{x_{1}x_{17}}
  \quad \Rightarrow~   
  x_{5} =1 - \frac{\bar{e}_{5}}{\bar{e}_{17}}\frac{x_{15}y_{17}}{z\,x_{1}x_{17}},
\end{equation}
\begin{equation}\label{eq:a23}
\frac{y_{3}}{y_{5}}=\frac{e_{3}}{e_{5}}\frac{1}{x_{5}}
  \quad \Rightarrow~   
  x_{3} =1- \frac{\bar{e}_{3}}{\bar{e}_{5}}\frac{y_{5}}{x_{5}},
\end{equation}
\begin{align}\label{eq:a24}
\frac{y_{9}}{y_{3}y_{8}}=\frac{e_{9}}{e_{3}e_{8}}\frac{x_{17}}{x_{3}x_{8}} 
\quad \Rightarrow~   
  x_{9} =1- \frac{\bar{e}_{9}}{\bar{e}_{3}\bar{e}_{8}}\frac{x_{17}y_{3}y_{8}}{x_{3}x_{8}},
\end{align}
\begin{equation}\label{eq:a25}
  \frac{y_{11}}{y_{9}}=\frac{1}{x_{9}}
 \quad \Rightarrow~    
  x_{11} = 2 - \frac{1}{x_{9}},
\end{equation}
\begin{equation}\label{eq:a26}
\frac{y_{13}}{y_{11}}=\frac{e_{13}}{e_{11}}\frac{1}{x_{11}}
 \quad \Rightarrow~    
  x_{13} = 1 - \frac{\bar{e}_{13}}{\bar{e}_{11}}\frac{y_{11}}{x_{11}},
\end{equation}
\begin{equation}\label{eq:a27}
  \frac{y_{14}}{y_{13}}=\frac{e_{14}}{e_{13}}\frac{1}{x_{9}x_{13}}
 \quad \Rightarrow~      
  x_{14} = 1 - \frac{\bar{e}_{14}}{\bar{e}_{13}}\frac{y_{13}}{x_{13}}.
\end{equation}
Substitution of the $x_m$ and $y_m$ from \eqref{eq:a17}-\eqref{eq:a27} into \eqref{eq:a16}
 yields the cubic equation,
\begin{equation}\label{eq:a1}
  1 - a_{1}\zeta -a_{2}\zeta^{2} + a_{3}\zeta^{3}=0,
\end{equation}
with coefficients,
\begin{align}\label{eq:a2}
  a_{1} & = 1+\bar{e}_{1}+2z\bar{e}_{15}, \nonumber \\
  a_{2} & = z^2[\bar{e}_{17}-\bar{e}_{15}^2] 
  +z[\bar{e}_{3}+2\bar{e}_{8}+\bar{e}_{5}-2(1+\bar{e}_{1})\bar{e}_{15} ],
  \nonumber \\ 
  a_{3} & = z^2[(1+\bar{e}_{1})(\bar{e}_{17}-\bar{e}_{15}^2) + (\bar{e}_{3}+2\bar{e}_{8}+\bar{e}_{5})e_{15} \nonumber \\ &\hspace{25mm} -\bar{e}_{14}-2\bar{e}_{9} -\bar{e}_{13}]. 
\end{align}
A unique physical solution is guaranteed if the modified energy parameters satisfy the inequalities,
\begin{equation}\label{eq:a6}
\sqrt{\bar{e}_{17}}>\bar{e}_{15},\quad
(\bar{e}_3+2\bar{e}_8+\bar{e}_5)\bar{e}_{15}
>\bar{e}_{14}+2\bar{e}_9+\bar{e}_{13},
\end{equation}
which are not restrictive from a physical perspective for the intended applications.
If the $\bar{e}_m$ satisfy the relation,
\begin{equation}\label{eq:a7}
(\bar{e}_3+2\bar{e}_8+\bar{e}_5)\sqrt{\bar{e}_{17}}
=\bar{e}_{14}+2\bar{e}_9+\bar{e}_{13},
\end{equation}
which is consistent with (\ref{eq:a6}), then the cubic equation (\ref{eq:a1}) factorizes with one particular root of the quadratic factor identifiable as the physically relevant solution.
The partition function in closed form for this case reads,
  \begin{align}\label{eq:a8}
    \bar{Z} &= \frac{1}{2}\Big[
    1+\bar{e}_{1}+(\bar{e}_{15}+\sqrt{\bar{e}_{17}})z \\ \nonumber  
    &+
    \sqrt{\big(1+\bar{e}_{1}-(\bar{e}_{15}+\sqrt{\bar{e}_{17}})z\big)^{2}
      -4z(\bar{e}_{3}+2\bar{e}_{8}+\bar{e}_{5})}\,\Big].
  \end{align}

From this result we infer explicit expressions for the variables $x_m$ via the Eqs.~(\ref{eq:a3})-(\ref{eq:a27}) and for the particle population densities $\bar{N}_m$ via the solution of the linear equations (\ref{eq:15}) with the $w_m$ from (\ref{eq:17}).
The excess volume of the jammed macrostate then follows from (\ref{eq:7})
and its entropy from (\ref{eq:20}).
  
\subsection{Default energy parameters}\label{sec:appa-1}
We continue with the default choice (\ref{eq:39}) for the $p_m$ that go into the energy parameters $\bar{e}_m$ as stated in (\ref{eq:a5}).
For the implementation of this choice, we introduce the non-negative quantities,
\begin{equation}\label{eq:a10}
 f_i\doteq e^{-\beta(\tilde{V}_i-\tilde{V}_\mathrm{b})}\quad i=\mathrm{a}, \mathrm{c}, \mathrm{d}, \mathrm{e}, \mathrm{f},
\end{equation}
associated with the volumes of two-disk tiles listed in Table ~\ref{tab:t2}. 
We can thus write,
\begin{align}\label{eq:a51}
& \bar{e}_1=f_\mathrm{d}/f_\mathrm{a},\quad
\bar{e}_3=f_\mathrm{c}^2/f_\mathrm{a}^2,\quad
\bar{e}_5=f_\mathrm{f}^2/f_\mathrm{a}^2, 
\nonumber \\
& \bar{e}_7=\bar{e}_8=f_\mathrm{c}f_\mathrm{f}/f_\mathrm{a}^2,\quad
\bar{e}_9=\bar{e}_{11}=f_\mathrm{c}f_\mathrm{f}/f_\mathrm{a}^3,
\nonumber \\
& \bar{e}_{13}=f_\mathrm{f}^2/f_\mathrm{a}^3,\quad
\bar{e}_{14}=f_\mathrm{c}^2/f_\mathrm{a}^3,
\nonumber \\
& \bar{e}_{15}=\bar{e}_{16}=f_\mathrm{e}/f_\mathrm{a},\quad
\bar{e}_{17}=1/f_\mathrm{a}^2.
\end{align}
The partition function (\ref{eq:a8}) turns into an explicit, parameterless  function of $\beta$ and $z$:
\begin{align}\label{eq:a11}
    \bar{Z}(\beta,z) &= \frac{1}{2f_\mathrm{a}}\Bigg[(1+f_\mathrm{e})z+f_\mathrm{a}+f_\mathrm{d} \\
    &-\sqrt{\Big((1+f_\mathrm{e})z-f_\mathrm{a}-f_\mathrm{d}\Big)^{2}
      +4z(f_\mathrm{c}+f_\mathrm{f})^{2}}\;\;\Bigg]. \nonumber
  \end{align}

Next we aim to establish the relation (\ref{eq:8}) between the chemical potential $\mu$, which enters $\bar{Z}$ via the fugacity $z$, and the (experimentally controllable) fraction $\bar{N}_\mathrm{S}$ of small disks.  
Toward this goal, a more compact rendering of the result (\ref{eq:a11}) is useful. 
We can write, 
\begin{equation}\label{eq:a12}
  \bar{Z}(\beta,z)
  = \frac{Z_{0}}{2}\left[1+
  \frac{z}{z_\mathrm{h}} +\sqrt{\left(1-\frac{z}{z_\mathrm{h}}\right)^{2}\!\!+4 \delta^2\,\frac{z}{z_\mathrm{h}}}\;\right].
\end{equation}
The significance of the three parameters, 
\begin{align}\label{eq:a13}
Z_{0} &\doteq \frac{f_\mathrm{a}+f_\mathrm{d}}{f_\mathrm{a}},\quad 
\delta(\beta) \doteq \frac{f_\mathrm{c}+f_\mathrm{f}}{\sqrt{(1+f_\mathrm{e})(f_\mathrm{a}+f_\mathrm{d})}}, \nonumber \\
z_\mathrm{h} &\doteq \frac{f_\mathrm{a}+f_\mathrm{d}}{1+f_\mathrm{e}} \quad :~
0<z_\mathrm{h}<\delta<1<Z_0<2,
\end{align}
will become clear shortly.
$\bar{N}_\mathrm{S}$ can be extracted from the partition function via derivatives,
\begin{equation}\label{eq:a14}
  \bar{N}_\mathrm{S} = \frac{1}{\beta \bar{Z}} \frac{\partial \bar{Z}}{\partial \mu}
  = \frac{z}{\bar{Z}} \frac{\partial \bar{Z}}{\partial z},
\end{equation}
to be carried out on expression (\ref{eq:a12}):
\begin{equation}\label{eq:a15}
  \bar{N}_\mathrm{S}(\beta,z) = \frac{1}{2}
  \left[1 +  \frac{z/z_\mathrm{h}-1}{\sqrt{(z/z_\mathrm{h}-1)^{2}+4\delta^2 z/z_\mathrm{h}}}\right].
\end{equation}
For given $\beta$ (contained in $\delta$ and $z_\mathrm{h}$), $\bar{N}_\mathrm{S}$ is a rather simple, monotonically increasing, function of $z$.
The limit ${\bar{N}_\mathrm{S}=0}$ ($\bar{N}_\mathrm{S}=1$) is realized for $z\to0$ ($z\to\infty$).
The case $\bar{N}_\mathrm{S}=\frac{1}{2}$ is realized for $z=z_\mathrm{h}$ (half and half).
For what follows we need the inverse relation $z(\beta,\bar{N}_\mathrm{S})$: 
\begin{align}\label{eq:a28}
\frac{z}{z_\mathrm{h}}
=\exp\Big(2\mathrm{Arsinh}(\eta_\mathrm{S}\delta)\Big),\quad
\eta_\mathrm{S}= \frac{\bar{N}_\mathrm{S}-\frac{1}{2}}{\sqrt{\bar{N}_\mathrm{S}\big(1-\bar{N}_\mathrm{S}\big)}}.
\end{align}
The antisymmetry of $\eta_\mathrm{S}$ about $\bar{N}_\mathrm{S}=\frac{1}{2}$ is another manifestation of symmetry. 
Expressions (\ref{eq:a3})-(\ref{eq:a27}) can now be simplified.
The solution of the linear Eqs.~(\ref{eq:15}) for the particle population densities then yield the compact expressions,
\begin{equation}\label{eq:a39}
\bar{N}_{1} = \bar{N}_{2} 
=\frac{y_{1}x_{15}x_{17}}{2D},\quad
\bar{N}_{3} = \bar{N}_{4} 
=\bar{N}_{1}\frac{y_{3}x_{1}x_{5}}{y_{1}},
 \end{equation}
\begin{equation}\label{eq:a40}
\bar{N}_{5} = \bar{N}_{6} 
=\bar{N}_{1}\frac{y_{5}x_{1}}{y_{1}},\quad
 \bar{N}_{7} = \frac{y_{7}\zeta x_{15}x_{17}}{x_{7}x_{8}D},
\end{equation}
\begin{equation}\label{eq:a41}
 \bar{N}_{8} = \bar{N}_{7}\frac{y_{8}x_{7}}{y_{7}},\quad
\bar{N}_{9} = \frac{y_{9}x_{1}x_{3}x_{5}x_{15}x_{17}}{2D}, 
\end{equation}
\begin{equation}\label{eq:a42}
\bar{N}_{11} = \bar{N}_{12}  
=\bar{N}_{9}\frac{y_{11}x_{9}}{y_{9}},\quad
\bar{N}_{13} = \bar{N}_{11}\frac{2y_{13}x_{11}}{y_{11}}, 
\end{equation}
\begin{equation}\label{eq:a43}
 \bar{N}_{14}  = \bar{N}_{13}\frac{y_{14}x_{13}}{y_{13}},\quad
 \bar{N}_{17} = \frac{y_{17}(x_{1}-\zeta)x_{15}}{D},
\end{equation}
\begin{align}\label{eq:a44}
 \bar{N}_{15} = \bar{N}_{16} 
  =\frac{y_{15}}{D}\Bigg[x_{1} &-\zeta\left(1+\frac{x_{17}y_{8}}{x_{8}}\right) \nonumber \\
 &  -\frac{x_{1}x_{17}(1-x_{3}x_{5})}{2}\Bigg], 
\end{align}
with $\zeta=1/\bar{Z}$ from (\ref{eq:a12}) and 
\begin{align}\label{eq:a45}
D &\doteq  (2-x_{17})(x_{1}-\zeta) \nonumber \\
&\hspace{10mm}+x_{17}\left(x_{15}+x_{1}x_{3}x_{5}
-\frac{\zeta}{x_{7}x_{8}}\right).
\end{align}
Excess volume (\ref{eq:7}) and entropy (\ref{eq:20}), which are functions of the $\bar{N}_m$ follow directly.

\subsection{Modified energy parameters}\label{sec:appa-2}
The results presented in Secs.~\ref{sec:ssm-b} and \ref{sec:ssm-c} require specific modifications in the energy parameters $\bar{e}_m$ as explained in Secs.~\ref{sec:jam-prot} and \ref{sec:ssm}.
All possible modifications are combinations of changes in the $\bar{e}_m$.
We noted that the criterion for distinct ordering tendencies hinges on whether the quantity $\Delta\mathcal{V}$ with default value (\ref{eq:31}) remains positive or goes negative.
The inequality $\Delta\mathcal{V}>0$ implies the inequality,
\begin{equation}\label{eq:a52}
\frac{f_\mathrm{c}^2}{f_\mathrm{a}}=e^{-\beta\Delta\mathcal{V}}<1,
\quad \Delta\mathcal{V}=0.19748\ldots,
\end{equation}
to be satisfied for the default energy parameters.
The numerical value for $\Delta\mathcal{V}$ stated in (\ref{eq:a52}) pertains to the disk diameters used throughout this study.
For the case described in Sec.~\ref{sec:ssm-b} we replace $\Delta\mathcal{V}$ by $-\Delta\mathcal{V}$ in (\ref{eq:a52}), which changes the ordering tendency.
The borderline case described in Sec.~\ref{sec:ssm-c} pertains to setting $\Delta\mathcal{V}=0$ in 
(\ref{eq:a52}).
In practice, we keep the default value of $f_\mathrm{a}$ and modify $f_\mathrm{c}$ in two out of three cases:
\begin{equation}\label{eq:a53}
f_\mathrm{c}=\left\{ \begin{array}{ll} \displaystyle
\sqrt{f_\mathrm{a}}e^{-\beta\Delta\mathcal{V}/2} & 
:~ \mathrm{Sec.~\ref{sec:ssm-a}},  \\ \rule[-2mm]{0mm}{7mm}
\displaystyle \sqrt{f_\mathrm{a}}e^{\beta\Delta\mathcal{V}/2} & 
:~ \mathrm{Sec.~\ref{sec:ssm-b}},  \\ \rule[-2mm]{0mm}{6mm}
\displaystyle \sqrt{f_\mathrm{a}} & 
:~ \mathrm{Sec.~\ref{sec:ssm-c}}. \end{array} \right.
\end{equation}

\subsection{Small-disk reference state}\label{sec:appa-3}
We stated in Sec.~\ref{sec:geom} that starting from reference state (\ref{eq:36}), representing the most compact configuration of small disks, instead of the large-disk configuration (\ref{eq:2}), makes no difference mathematically and produces the same physics.
The switch amounts to interchanging two pairs of the six tile volumes listed in Table~\ref{tab:t2}:
\begin{equation}\label{eq:a54}
\tilde{V}_\mathrm{a} \longleftrightarrow \tilde{V}_\mathrm{b},\quad 
\tilde{V}_\mathrm{d} \longleftrightarrow \tilde{V}_\mathrm{c}.
\end{equation}
In the statistical mechanical analysis carried out above (Sec~\ref{sec:appa-1}), the switch entails the following substitutions:
\begin{equation}\label{eq:a55}
\bar{N}_\mathrm{L} \longleftrightarrow \bar{N}_\mathrm{S},\quad 
z_\mathrm{h} \longleftrightarrow z_\mathrm{h}^{-1},\quad
z \longleftrightarrow z^{-1}.
\end{equation}
The mathematical simplicity associated with the switch of reference state is owed to the symmetry of the model.

\subsection{Entropy and excess volume}\label{sec:appa-4}
For the graphical representations in Sec.~\ref{sec:scen1} of excess volume $\bar{V}$ we use (\ref{eq:7}) with the $\bar{N}_m$ as determined analytically in the preceding parts of this appendix. We could proceed likewise for the configurational entropy $\bar{S}$ by using the same $\bar{N}_m$ in (\ref{eq:20}).
However, it is possible derive a closed-form expression for $\bar{S}$ as a function of $\bar{N}_\mathrm{S}$, $\bar{N}_\mathrm{L}$, and $\beta$ from the partition function (\ref{eq:a12}) via 
\begin{align}\label{eq:a60}
\bar{S}=\beta^2\frac{\partial\bar{\Omega}}{\partial\beta},\quad
\bar{\Omega}=-\frac{1}{\beta}\ln\bar{Z},
\end{align}
with use of (\ref{eq:a15}):
\begin{align}\label{eq:a56}
\bar{S} &=\bar{N}_\mathrm{L}\bar{S}_\mathrm{S}^{(0)}
+\bar{N}_\mathrm{S}\bar{S}_\mathrm{L}^{(0)}
+\frac{1}{2}\ln\big(1-\delta^2\big) \\
&+\mathrm{Arsinh}\left(\frac{\eta_\mathrm{S}\delta}{(\bar{N}_\mathrm{S}-\bar{N}_\mathrm{L})\sqrt{1-\delta^2}}\right) \nonumber \\
&-(\bar{N}_\mathrm{S}-\bar{N}_\mathrm{L})\mathrm{Arsinh}(\eta_\mathrm{S}\delta)
  \nonumber \\
&+\frac{\beta\nu\delta\sqrt{\bar{N}_\mathrm{S}\bar{N}_\mathrm{L}}}
{\sqrt{1-\delta^2}}\,\exp\left(-\mathrm{Arsinh}\left(\frac{\eta_\mathrm{S}\delta}{(\bar{N}_\mathrm{S}-\bar{N}_\mathrm{L})\sqrt{1-\delta^2}}\right)\right).
\nonumber
\end{align}
where
\begin{align}\label{eq:a57}
\bar{S}_\mathrm{S}^{(0)}=\ln\left(1+\frac{f_\mathrm{d}}{f_\mathrm{a}}\right)
+\frac{\ln(f_\mathrm{a}/f_\mathrm{d})}{1+f_\mathrm{a}/f_\mathrm{d}},
\end{align}
\begin{align}\label{eq:a58}
\bar{S}_\mathrm{L}^{(0)}=\ln\left(1+f_\mathrm{e}\right)
+\frac{\ln(1/f_\mathrm{e})}{1+1/f_\mathrm{e}},
\end{align}
\begin{align}\label{eq:a59}
\beta\nu(\beta) &=\frac{f_\mathrm{a}\ln f_\mathrm{a}+f_\mathrm{d}\ln f_\mathrm{d}}
{f_\mathrm{a}+f_\mathrm{d}}
+\frac{f_\mathrm{e}\ln f_\mathrm{e}}{1+f_\mathrm{e}} \nonumber \\
&-2\frac{f_\mathrm{c}\ln f_\mathrm{c}+f_\mathrm{f}\ln f_\mathrm{f}}
{f_\mathrm{c}+f_\mathrm{f}},
\end{align}
$\delta(\beta)$ from (\ref{eq:a13}) and $\eta_\mathrm{S}$ from (\ref{eq:a28}).

%
\section{MODEL OF HIGHER STABILITY}\label{sec:appb}
%

We begin with combinatorial specifications after the modifications discussed in Sec.~\ref{sec:scen2} and the mergers carried in its wake.
They are compiled in Table~\ref{tab:t6}.
The partition-function again reduces to a function of $x_{1}$.
The reasoning behind this conclusion is the same as in Appendix~\ref{sec:appa}, but the result is different:
\begin{equation}\label{eq:B1}
  \bar{Z} = \frac{\bar{e}_{1}x_{1}}{1-x_{1}}.
\end{equation}

With $\zeta=\bar{Z}^{-1}$ and $y_m\doteq 1-x_m$ we can express the solutions of Eqs.~(\ref{eq:19})
in succession as follows:
\begin{align}\label{eq:B2}
  x_{1} & = 1-y_{1},\quad  y_{1} = \frac{f_\mathrm{d}\zeta}{f_\mathrm{a}+f_\mathrm{d}\zeta},
  \\ \label{eq:B3}
  x_{15} & = 1 - y_{15},\quad
  y_{15} =\frac{f_\mathrm{e}z\zeta}{f_\mathrm{a}+f_\mathrm{e}z\zeta},
  \\ \label{eq:B4}
  x_{17} & = 1 - y_{17},\quad y_{17}=(1+z\zeta f_\mathrm{e}/f_\mathrm{a})^{2}z^{2}\zeta^{2}/f_\mathrm{a}^{2},
  \\ \label{eq:B5}
  x_{8} & = 1 -y_{8}, \quad  y_{8}= \frac{f_\mathrm{c}f_\mathrm{f}\,y_{15}}{f_\mathrm{a}f_\mathrm{e}\,x_{15}x_{17}+f_\mathrm{c}f_\mathrm{f}\,y_{15}},
  \\ \label{eq:B6}
  x_{7} & = 1- y_{7},\quad  y_{7} = \frac{y_{8}}{2-x_{8}},
  \\ \label{eq:B7}
  x_{3} & = 1-y_{3},\quad y_{3}= \frac{f_\mathrm{c}^{2}}{z}\frac{x_{15}y_{17}}{x_{1}x_{17}},
  \\ \label{eq:B8}
  x_{9} &= 1 - y_{9}, \quad
  y_{9}=\frac{f_\mathrm{f}}{f_\mathrm{d}f_\mathrm{c}}z\frac{y_{1}y_{3}}{x_{1}x_{3}},
  \\ \label{eq:B9}
  x_{11} & = 1- y_{11}, \quad y_{11} = \frac{y_{9}}{x_{9}},
  \\ \label{eq:B10}
  x_{13}  & = 1- y_{13},\quad y_{13}=\frac{f_\mathrm{f}}{f_\mathrm{c}} \frac{y_{11}}{x_{11}},
  \\ \label{eq:B11}
  x_{14} & = 1-y_{14}, \quad y_{14}=\frac{f_\mathrm{c}^{2}}{f_\mathrm{f}^{2}} \frac{y_{13}}{x_{13}}.
\end{align}

\begin{table}[htb]
  \caption{Scaled capacity constants $\bar{A}_m$, quantum numbers $s_m$, and interaction coefficients $\hat{g}_{mn}$ of the particle species after merging operations. The $\Delta V_m$ are as in Table~\ref{tab:t8}.}\label{tab:t6}
\begin{center}
\begin{tabular}{r|rr}
$m$ & $\bar{A}_m$ & $s_m$\\ \hline
1 & 1 & 0 \\
3 & 1 & 1 \\
7 & 1 & 1 \\
8 & 1 & 1 \\
9 & 1 & 2 \\
11 & 1 & 2 \\
13 & 1 & 2 \\
14 & 1 & 2 \\
15 & 0 & 1 \\
17 & 0 & 2 \\
\end{tabular} \hspace*{5mm}
\begin{tabular}{r|rrrrrrrrrr} 
  $\hat{g}_{mn}$
    &~1 &~3 &~7 &~8 &~9 &11& 13& 14& 15& 17\\ \hline
  1 & 2 & 1 & 1 & 1 & 2 & 2 & 2 & 2 & 1 & 2\\ 
  3 & 1 & 2 & 1 & 1 & 2 & 2 & 2 & 2 & 1 & 2\\
  7 & 1 & 2 & 2 & 1 & 3 & 3 & 3 & 3 & 1 & 2\\
  8 & 1 & 2 & 2 & 2 & 3 & 3 & 3 & 3 & 1 & 2\\
  9 & 1 & 2 & 1 & 1 & 3 & 2 & 2 & 2 & 1 & 2\\
  11& 1 & 2 & 1 & 1 & 3 & 3 & 2 & 2 & 1 & 2\\
  13& 1 & 2 & 1 & 1 & 3 & 3 & 3 & 2 & 1 & 2\\   
  14& 1 & 2 & 1 & 1 & 3 & 3 & 3 & 3 & 1 & 2\\
  15& 0 &-1 & 0 & 0 &-1 &-1 &-1 &-1 & 1 &-2\\   
  17& 0 &-1 &-1 &-1 &-1 &-1 &-1 &-1 & 0 & 0\\   
\end{tabular}
\end{center}
\end{table} 

Eliminating the $x_m$ and $y_m$ from \eqref{eq:B1}-\eqref{eq:B11} leads to a polynomial equation of order 6 for $\zeta$:
\begin{subequations}\label{eq:B12}
  \begin{align}\label{eq:B12a}
    p_{z}(\zeta)
    & = \big(1-(1+z\zeta f_\mathrm{e}/f_\mathrm{a})^{2}z^{2}
    \zeta^{2}/f_\mathrm{a}^{2}\big)p_{0}(\zeta)
    \\ \nonumber 
    & -(1+\zeta f_\mathrm{d}/f_\mathrm{a})\big(
      (f_\mathrm{c}+2f_\mathrm{f})f_\mathrm{c}f_\mathrm{a}\\  \nonumber
      &+(f_\mathrm{c}
      +f_\mathrm{f})^{2}(1+z\zeta f_\mathrm{e}/f_\mathrm{a})z\zeta
      +z\zeta f_\mathrm{c}^{2}f_\mathrm{e}\big)z\zeta^{2}/f_\mathrm{a}^{3}
  \end{align}
  \begin{equation}\label{eq:B12b}
   p_{0}(\zeta)\doteq 1-\zeta-f_\mathrm{d}/f_\mathrm{a}\zeta^{2},
  \end{equation}
\end{subequations}
In expanded form, this polynomial reads
\begin{align}\label{eq:B13}
  p_z(\zeta) & = \sum_{\ell=0}^{6}a_{\ell}\zeta^{\ell},
  \end{align}
  with coefficients
  \begin{subequations}\label{eq:B14}
    \begin{align}\label{eq:B14a}
      a_{0} & = 1,
      \\ \label{eq:B14b}
      a_{1} &= -1,\\ \label{eq:B14c}
      a_{2} &= -\big(f_\mathrm{d}f_\mathrm{a}+(f_\mathrm{c}+2 f_\mathrm{f})f_\mathrm{c}z +z^{2}\big)/f_{a}^{2} < 0, \\ \label{eq:B14d}
      a_{3} &= -\big((f_\mathrm{c}+2 f_\mathrm{f})f_\mathrm{c}f_\mathrm{d}z \\\nonumber 
              &+\big((f_\mathrm{c}+f_\mathrm{f})^{2} -f_\mathrm{a} +f_\mathrm{c}^{2} f_\mathrm{e}\big) z^{2}
              +2f_\mathrm{e}z^{3}\big)/f_\mathrm{a}^{3} \leq 0,
      \\ \label{eq:B14e}
      a_{4} &= -\Big(\big((f_\mathrm{c}+f_\mathrm{f})^{2}-f_\mathrm{a}+f_\mathrm{c}^{2}f_\mathrm{e}\big)f_\mathrm{d}z^{2}
      \nonumber \\
              &+\big((f_\mathrm{c}+f_\mathrm{f})^{2}-2f_\mathrm{a}\big)f_\mathrm{e} z^{3}
              +f_\mathrm{e}^{2}z^{4}\Big)/f_\mathrm{a}^{4}  \leq 0,
      \\ \label{eq:B14f}
      a_{5} &= -\Big(\big((f_\mathrm{c}+f_\mathrm{f})^{2}-2f_\mathrm{a}\big)f_\mathrm{d}f_\mathrm{e} z^{3}-f_\mathrm{e}^{2}z^{4}\Big)/f_\mathrm{a}^{4}, 
      \\ \label{eq:B14g}
      a_{6} &= f_\mathrm{e}^{2}f_\mathrm{d}z^{4}/f_\mathrm{a}^{5} \geq 0.
    \end{align}
\end{subequations}
It can be shown that a unique positive physical root exists and has the range,
\begin{align}\label{eq:B15}
0 < \zeta <  \min\Big\{1,\frac{f_\mathrm{a}}{z}
\frac{2}{\sqrt{1+4f_\mathrm{e}}+1}\Big\}.
\end{align}

To derive the asymptotic expansion for large $\beta$ (implying small $f_\mathrm{a}/z$ with $z>0$), we consider the polynomial equation $p(\zeta_{*})=0$ with ingredients in (\ref{eq:B13}) and (\ref{eq:B14}). 
Analyzing the terms for each power of $\zeta_*$ individually, we find that $a_0\zeta_*$ and $a_2\zeta_*^2$ are of $\mathcal{O}(1)$ and $a_i\zeta_*^i$ for $i=1,3,4,5,6$ are of $\mathcal{O}(f_\mathrm{a}/z)$.
The solution to leading order becomes
\begin{align}\label{eq:B16}
  \zeta_{*} =  \frac{1}{\sqrt{a_{2}}} \leadsto \frac{f_\mathrm{a}}{z}.
\end{align}
A Newton iteration,
\begin{align}\label{eq:B17}
  \zeta_{*}^{(i+1)} = \zeta_{*}^{(i)} - \frac{p_{z}(\zeta_{*}^{(i)})}{p_{z}'(\zeta_{*}^{(i)})},\quad \zeta^{(0)}=\frac{f_\mathrm{a}}{z},
\end{align}
leads to a consistent expansion after just two iterations:
\begin{align}\label{eq:B18}
  \zeta_{*} =\frac{f_\mathrm{a}}{z}(1-f_\mathrm{e}+ ...).
\end{align}
In the mapping from the segregation to the alternation regime as distinguished in  (\ref{eq:a53}), which implies the substitution, $f_\mathrm{c} \mapsto f_\mathrm{a}/f_\mathrm{c}$, we again find (\ref{eq:B18}).
The exact solution worked out in Appendix~\ref{sec:appa} produces the exact same first two terms in the corresponding expansion.
Thus, the asymptotic behavior of $\zeta_{*}$ is consistent across both models -- the more symmetric one and the more stable one -- up to second order.



\begin{thebibliography}{100}
\bibitem{note2}
Technically, the activation of one host particle 3 or 14 needs to be activated in order to accommodate tags 15 and 17.
In the thermodynamic limit (\ref{eq:13}), that host leaves no trace in any quantity of interest.
\end{thebibliography}
\end{document}